\documentclass[11pt,english]{article}
\usepackage{geometry}
\usepackage{setspace}
\usepackage[T1]{fontenc}
\usepackage[utf8]{inputenc}
\usepackage{babel}
\usepackage{varioref}
\usepackage{amsmath,nccmath}
\numberwithin{equation}{section}
\usepackage{amssymb}
\usepackage{mathrsfs} 
\usepackage{graphicx}
\usepackage{floatrow}
\usepackage[labelfont=bf]{caption}
\usepackage{appendix}
\usepackage{slashed}
\usepackage{svg}
\usepackage{bm}
\usepackage [autostyle, english = american]{csquotes}
\MakeOuterQuote{"}
\makeatletter
\usepackage{enumitem}
\usepackage{hyperref}
\usepackage{xcolor}
\usepackage{authblk}
\usepackage{amsthm}
\usepackage{graphicx}
\usepackage{hyperref}
\hypersetup{
    colorlinks,
    linkcolor={blue!60!black},
    citecolor={green!60!black},
    urlcolor={blue!80!black}
}
\usepackage{mathtools}
\usepackage{mathrsfs}
\usepackage{comment}
\usepackage{todonotes}
\allowdisplaybreaks
\newtheorem{theorem}{Theorem}[section]

\newtheorem{lemma}{Lemma}[section]
\newtheorem{definition}{Definition}[section]

\theoremstyle{remark}

\theoremstyle{plain}

\newenvironment{nalign}{
    \begin{equation}
    \begin{aligned}
}{
    \end{aligned}
    \end{equation}
    \ignorespacesafterend
}
\theoremstyle{remark}

\newcommand{\dd}{\mathop{}\!\mathrm{d}}
\renewcommand{\O}{\mathcal{O}}
\newcommand{\e}{\mathrm{e}}
\newcommand*{\defeq}{\mathrel{\rlap{%
                     \raisebox{0.3ex}{$\m@th\cdot$}}%
                     \raisebox{-0.3ex}{$\m@th\cdot$}}%
                     =}

\newcommand{\overone}{\stackrel{\mbox{\scalebox{0.4}{(1)}}}}

\newcommand{\lap}{\mathring{\slashed{\Delta}}}

\newcommand{\tr}{\mathrm{tr}}
\newcommand{\pu}{\partial_u}
\newcommand{\pv}{\partial_v}

\newcommand{\gs}{\overone{\slashed{g}}}
\newcommand{\gsh}{\overone{\hat{\slashed{g}}}}
\newcommand{\trg}{{\tr{\gs}}}
\newcommand{\Om}{\overone{\Omega}}

\newcommand{\trx}{\overone{(\Omega \tr\chi)}}
\newcommand{\trxb}{\overone{(\Omega \tr\underline\chi)}}
\newcommand{\xh}{\overone{\hat{\chi}}}
\newcommand{\xhb}{\overone{\hat{\underline\chi}}}

\newcommand{\alpas}[1][s]{\alpha^{[#1]}}

\newcommand{\be}{\overone{\beta}}

\newcommand{\K}{\overone{K}}
\newcommand{\Ps}{\overone{\Psi}}

\newcommand{\C}{\mathcal{C}}

\newcommand{\Cinn}{\mathcal{C}}
\newcommand{\Sone}{\mathcal{S}_1}
\newcommand{\Sinfty}{\mathcal{S}_{\infty}}
\newcommand{\Stwo}{\mathbb{S}^2}
\newcommand{\Scrim}{\mathcal{I}^{-}}
\newcommand{\Scrip}{\mathcal{I}^{+}}

\newcommand{\aldata}{\mathscr{A}}
\newcommand{\albdata}{\underline{\mathscr{B}}}

\title{The Case Against Smooth Null Infinity IV:\\Linearised Gravity Around Schwarzschild -- An Overview} 

\author[1]{Lionor Kehrberger\thanks{kehrberger@mis.mpg.de}}
\affil[1]{University of Cambridge, Department of Applied Mathematics and Theoretical Physics, Wilberforce~Road, Cambridge~CB3~0WA, United Kingdom}
\date{April 30, 2024} 

\makeatother
\begin{document}
\pagenumbering{roman}

\maketitle 
\begin{abstract}
This paper is the fourth in a series dedicated to the mathematically rigorous asymptotic analysis of gravitational radiation under astrophysically realistic setups.  
It provides an overview of the physical ideas involved in setting up the mathematical problem, the mathematical challenges that need to be overcome once the problem is posed, as well as the main new results we obtain in the companion paper \cite{V}.

From the physical perspective, this includes a discussion of how Post-Newtonian theory provides a prediction on  the gravitational radiation emitted by $N$ infalling masses from the infinite past \textit{in the intermediate zone}, i.e.~up to some finite advanced time. 

From the mathematical perspective, we then take this prediction, together with the condition that there be no incoming radiation from $\mathcal{I}^-$, as a starting point to set up a scattering problem for the linearised Einstein vacuum equations around Schwarzschild and near spacelike infinity, and we outline how to solve this scattering problem and obtain the asymptotic properties of the scattering solution near $i^0$ and $\mathcal{I}^+$.

The full mathematical details are presented in the companion paper \cite{V}.

\end{abstract}
    \begingroup
\hypersetup{linkcolor=black}
    \tableofcontents{}
    \endgroup
\newpage
\pagenumbering{arabic}
\section{Introduction}
In this paper, we physically motivate and give an overview of mathematical constructions of solutions to the linearised Einstein vacuum equations around Schwarzschild with the following features:

\maketitle

\begin{enumerate}[label=(\roman*),leftmargin=*]
\item They describe the far field region of a system of $N$ infalling masses coming from the infinite past and following approximately hyperbolic Keplerian orbits.\label{list1}
\item They have no incoming radiation from past null infinity.\label{list2}
\item They violate peeling near past null infinity: Near $\Scrim$, the extremal component of the Weyl tensor $\Psi_4$ (a.k.a.~$\alpas[-2]$) decays like $r^{-4}$ rather than $r^{-5}$.\label{list3}
\item They violate peeling near future null infinity:
Near $\Scrip$, the other extremal component of the Weyl tensor, $\Psi_0$ (a.k.a.~$\alpas[+2]$), decays like $r^{-4}+r^{-5}\log^2 r$ rather than $r^{-5}$. 
In particular, the constructed class of spacetimes does not admit a smooth null infinity/conformal compactification.\label{list4}
\item They decay slower towards spacelike infinity than assumed in most stability works. Both $\Psi_0$ and $\Psi_4$ decay like $r^{-3}$ along $t=0$, as opposed to the $o(r^{-7/2})$-decay rate assumed in \cite{CK93}.\label{list5}
\end{enumerate}
In the list above, the bullet points~\ref{list2} and~\ref{list3} play the role of \textit{scattering data assumptions}, from which we will construct a (unique) \textit{scattering solution}. 
These scattering data assumptions, in turn, are motivated by the physical considerations involved in~\ref{list1}.
Finally,~\ref{list4} and~\ref{list5} play the role of theorems that we prove as we analyse asymptotic properties of this scattering solution.

The detailed construction and asymptotic analysis of these solutions are presented in the companion paper~\cite{V}. 
The purpose of the present paper 
is to already give a rough outline of the main problems, ideas and results of the construction, as well as a discussion of the physical motivation for the construction. 

The remainder of this introduction will provide some background and context for the general problem studied in the series of papers \textit{The Case Against Smooth Null Infinity}.
\subsection{Isolated systems}
The desire to study systems in isolation is ubiquitous in physics: 
Given a physical process, we first identify which parts of its surroundings and interactions we can neglect. We then seek a mathematical model that treats the process as independent of those surroundings and interactions. Finally, we attempt to study the resulting model---the \textit{isolated system}---using suitable mathematical methods, hoping that its predictions will accurately approximate the actual physical process.

Consider, for instance, the motion of Earth and Sun around each other within Newtonian theory. A simple way to study this in isolation is to disregard all other matter in the universe, i.e.~to model the surroundings of Earth and Sun by \textit{vacuum}, and to only consider (Newtonian) gravitational interactions between Earth and Sun. 
By taking into account further interactions with other planets etc., the resulting predictions can be made more and more accurate; but the conceptual approach to \textit{isolate} the system remains the same.

Two reasons why this conceptual approach is simple within Newtonian theory are that Newtonian theory assumes that matter moves on a fixed geometric background, and that the theory has a simple and unambiguous realisation of vacuum. 

In General Relativity, however, the situation is different. The Einstein equations,
\begin{equation}
R_{\mu\nu}-\frac12 R g_{\mu\nu}=8\pi T_{\mu\nu}+\Lambda g_{\mu\nu},
\end{equation}
are equations for the geometry of an a priori unknown spacetime, and, even when we set the matter part as well as the cosmological constant to 0 (so as to ignore cosmological effects), the remaining Einstein vacuum equations,
\begin{equation}\label{EVE}
R_{\mu\nu}=0,
\end{equation}
still form a highly complicated system of geometric PDE with a large space of solutions, the dynamical degrees of freedom of these solutions being carried by gravitational radiation which, if focussed enough, can even create black holes \cite{Chr09}. 
Intuitively, we want to exclude such events from occurring and expect that the gravitational radiation of isolated systems should disperse at large distances; we want the spacetime to asymptote to the Minkowski spacetime in some very weak sense. Making this precise requires a thorough understanding of the asymptotic structure of gravitational radiation, and, indeed, notions of isolated systems in GR are often masked behind expressions such as ``asymptotic flatness''.

In the next two subsections, we will give an overview of (and some commentary on) some of the historically most influential works and approaches to this problem, and explain some short-comings of these approaches. 
These approaches model isolated systems based on what can be called \textit{the asymptotic problem}. 
We will then go over certain arguments against these approaches, which also take into account what can be called the \textit{generation problem} and the \textit{propagation problem} of gravitational radiation.

We will also argue that the matter of how to model isolated systems is not only relevant from an epistemological point of view, but that it has very concrete implications for important mathematical problems and even astrophysical effects such as gravitational wave tails.

\subsubsection{The works of Bondi, Sachs et al., and the peeling property}\label{sec:intro:BS}
In a series of papers by Bondi, van der Burg, Metzner and Sachs \cite{Bondi60Nature,SeriesVI,SeriesVII,SeriesVIII}, the authors introduced various ideas whose influence on future works concerning gravitational radiation is difficult to overestimate.
While the initial project had the aim of understanding conditions that guarantee that a given asymptotic region of spacetime only features outgoing radiation, 
 its scope quickly widened:
The authors essentially opened up a new field of study, centred around the question: To what degree can we understand aspects of spacetime by only analysing its possible asymptotic behaviour, i.e.~by its behaviour near infinity? 
For instance, can we deduce from soft arguments certain symmetries and regularity properties that gravitational radiation or spacetimes should satisfy, without studying the actual ``interior'' of spacetime? And can this potentially give us insights \textit{a posteriori} into the interior of spacetimes?
(Note that several modern approaches to quantum gravity ask exactly this question, with countless conjectured correspondences between the bulk and the boundary of spacetime.)

To go back to the concrete question at hand, the authors argued in \cite{SeriesVI,SeriesVII} that the gravitational field of (vacuum) spacetimes with no incoming radiation should, by analogy to the behaviour of linearised gravity around Minkowski (cf.~\S\ref{sec:phys}), admit power series expansions in $1/r$ along \textit{outgoing} null geodesics, $r$ denoting the luminosity parameter along these geodesics.
Operating under this assumption, a system of coordinates (``Bondi coordinates'') was constructed, and it was cautiously suggested in \cite{Sachs62BMS} that the notion of an isolated system/asymptotic flatness might be captured by the existence of such Bondi coordinates.
It was further shown that the Weyl curvature tensor ``peels'' towards $\Scrip$ \cite{SeriesVI,NP62Approach,SeriesVIII,GoldbergKerr64}, a statement which in Newman--Penrose notation can be written as
\begin{equation}\label{eq:intro:peeling}
\Psi_i=\Psi_i^0 r^{-5+i}+\mathcal{O}(r^{-6+i}),\quad i=0,\dots,4,
\end{equation}
where $\Psi_i$ denote the Newman--Penrose scalars \cite{NP62Approach} ($\Psi_0$ and $\Psi_4$ are defined in eq.~\eqref{eq:phys:defPsi0Psi4} of the present paper), and $\Psi_i^0$ are some functions independent of $r$.\footnote{For an overview over the field before the advent of the ideas discussed in the present section, we refer to the lecture notes \cite{TrautmanLN58} (from 1958) and references therein.}

At the technical level, this peeling property greatly simplifies the Einstein equations, transforming them into a set of a few relatively simple evolutionary equations and a set of algebraic identities.
Its imposition therefore led to a stark increase in activity and progress in studying the previously almost impenetrable Einstein equations.

At the physical level, however, there was debate from early on whether the peeling property captures the no incoming radiation condition, or whether it is even logically related to it (it only is for gravity linearised around the Minkowski spacetime), and several works \cite{CT72,NG82Nonpeeling,Winicour85LogarithmicAF,SeriesXIV} considered more general expansions of the Weyl curvature tensor than \eqref{eq:intro:peeling}. 
The approach of \textit{imposing certain asymptotic behaviour} and then studying the consequences, however, remained popular ever since. 
Furthermore, many of the early discoveries that came out of the general framework set up by Bondi et al., such as the Bondi mass loss formula \cite{Bondi60Nature}, or the BMS symmetry group at infinity~\cite{SeriesVII,SeriesVIII,Sachs62BMS,NP66Note}, were soon understood to be somewhat robust with respect to a broad class of violations of peeling.
\subsubsection{Penrose's smooth conformal compactification}\label{sec:intro:CC}
Only very briefly after the appearance of \cite{SeriesVI}, a concept of pristine geometric elegance and clarity that, in particular, captures the peeling property discussed above was introduced by Penrose \cite{Penrose63,Penrose65}. 
This concept gave a new, very concrete proposal for modelling isolated systems; we paraphrase it here:

An isolated astrophysical system should asymptotically become empty and approach the Minkowski spacetime.
The Minkowski spacetime has the property that it can be conformally compactified and, upon compactification, admits smooth conformal null boundaries called future and past null infinity ($\mathcal I^+$ and $\mathcal I^-$), respectively.
The hope is now that spacetimes describing isolated astrophysical processes should share this property. 
Precisely this idea is captured by the definition of \textit{asymptotic simplicity}:
Without going into details, a physical spacetime is called asymptotically simple if there exists another (unphysical) manifold and a conformal factor such that the conformally rescaled spacetime isometrically embeds into the unphysical manifold, and can be smoothly extended to its null boundaries. 
This property is also referred to as the spacetime possessing \textit{a smooth null infinity}.

Owing to the conformal properties of the Weyl tensor, spacetimes with a smooth null infinity satisfy the peeling property \eqref{eq:intro:peeling} near $\Scrip$. Similarly, along ingoing null geodesics near $\Scrim$, they satisfy
\begin{equation}\label{eq:intro:peelinpast}
\Psi_{4-i}=\Psi_i^{4-i,\mathrm{past}}r^{-5+i}+\mathcal{O}(r^{-6+i}),\quad i=0,\dots,4.
\end{equation}
Note that the original argument relating \eqref{eq:intro:peeling} to the absence of incoming radiation would similarly relate \eqref{eq:intro:peelinpast} to the absence of outoing radiation.

Since asymptotic simplicity defines a class of \textit{global} spacetimes, it has played a crucial role in understanding various global aspects of spacetimes, for instance related to black holes (which in most textbooks are defined using asymptotic simplicity).\footnote{We note in passing that, owing to another condition in the definition of asymptotic simplicity, not even the Schwarzschild spacetime is asymptotically simple due the presence of trapped null geodesics, but this is of no further relevance here.}

Again, the conceptual approach of asymptotic simplicity is to understand aspects of spacetime by certain considerations concerning their asymptotic behaviour, which in turn are motivated by soft/formal considerations, that is to say, considerations that do not directly take into account the physical structure of the system under consideration.

\begin{figure}[htpb]
\floatbox[{\capbeside\thisfloatsetup{capbesideposition={right,bottom},capbesidewidth=6cm}}]{figure}[\FBwidth]
{\caption{Depiction of a ``Bondi bomb'' spacetime, i.e.~a spacetime that is stationary in the past. The often made assumption of compactly supported initial data corresponds to precisely such a scenario. }\label{fig:BB}}
 {\includegraphics[width=90pt]{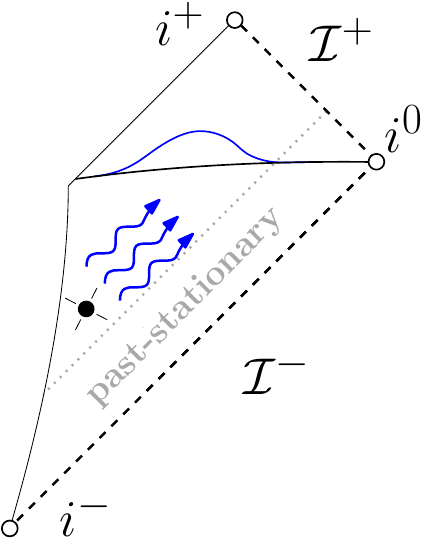}}
\end{figure}

Nonetheless, even though there was not yet a direct connection established between the generation of gravitational waves by some physical system and their asymptotic behaviour as posited by peeling or asymptotic simplicity, and even though the conformal irregularity of spatial infinity in the presence of non-trivial ADM mass was later understood to cause trouble to the simultaneous regularity of \textit{both} future \textit{and} past null infinity \cite{SchmidtStewart79,PorrillStewart81}, there was still a class of physical spacetimes for which it was relatively clear that they would possess a smooth null infinity, namely \textit{past-stationary spacetimes} (cf.~Fig.~\ref{fig:BB}). 
 Such spacetimes, sometimes referred to as \textit{Bondi bomb spacetimes}, are modelled to be exactly stationary up until some sudden, violent explosion of gravitational waves (which is triggered by something so weak that it is neglected within the model). 
It is relatively straight-forward to see that such a spacetime satisfies peeling at least until the retarded time at which the bomb explodes. Furthermore, by propagation of regularity arguments, the spacetimes will then continue to satisfy peeling for any finite retarded time. See \cite{Blanchet87, Friedrich86}.

From the mathematical perspective, this Bondi bomb scenario corresponds to the study of \textit{forward} Cauchy problems where the initial data are assumed to be compactly supported. (Notice that Bondi bomb spacetimes as depicted in~Fig.~\ref{fig:BB}, evolving from stationary to nonstationary spacetimes cannot arise from backwards evolution from Cauchy data for standard matter models; one would have to manually insert the explosion of the bomb into the matter model!)

\subsubsection{The case against smooth conformal compactification}\label{sec:intro:TCASNI}
The previous two subsections discussed the \textit{asymptotic problem of gravitational radiation} on relatively abstract grounds. 
We will now, in addition, take into account the \textit{generation} (what is the structure of gravitational radiation \textit{generated} by a physical process?) and the \textit{propagation problem} (how does this radiation \textit{propagate} through spacetime?) of gravitational radiation.
Of course, as long as the generation problem is studied at \textit{finite} time, one inevitably has to make \textit{a choice} for the asymptotic behaviour of radiation along that time slice.
On the other hand, if we study the generation and propagation problem in the \textit{infinite past}, we will obtain a dynamical prediction on the asymptotic behaviour of gravitational radiation! 

One of the most basic physical systems that any notion of isolated system should be able to describe is the two-body problem. 
Let us, for instance, think about two masses approaching each other from the infinite past. 
If these masses move at non-relativistic relative speeds, then Newtonian theory provides a reasonable approximation for the movement of these masses.
For similar reasons that the hydrogen atom is classically unstable, we are then forced to consider orbits which are unbound in the infinite past, i.e.~either approximately parabolic or hyperbolic Keplerian orbits (see \cite{WalkerWill79I}). 

Next, given this setup, we can attempt to understand the generation of gravitational radiation by these masses coming in from the infinite past using Post-Newtonian approximations for gravity linearised around Minkowski.
It then turns out, as was already understood very early on \cite{BardeenPress73,WalkerWill79}, that the quadrupolar radiation generated in this way fails to satisfy peeling near $\Scrim$ (eq.~\eqref{eq:intro:peelinpast}): 
The study of the \textit{generation problem} shows that the asymptotic structure of gravitational radiation as posited by peeling or asymptotic simplicity fails, and it \textit{constructively provides a different prediction for this asymptotic structure}. A brief reconstruction of this argument along with a generalisation to higher multipoles is provided in~\S\ref{sec:phys} of the present paper.

However, the same early arguments, by extending the assumed validity of the Post-Minkowskian predictions all the way to infinite advanced times, still found that the future null infinity of such spacetimes should be smooth, as they were still made within the framework of linearised gravity around Minkowski.
The general idea that peeling fails near $\Scrim$ but holds true near $\Scrip$ was somewhat substantiated by the evidence of \cite{SchmidtStewart79,PorrillStewart81} concerning the incompatibility between peeling at $\Scrim$ and at $\Scrip$.

New light on the matter was shed by Damour in 1986 \cite{Damour86} (see also \cite{Winicour84Extension}):
 In addition to taking into account the generation problem (to find that peeling fails near past null infinity), Damour also analysed the \textit{propagation problem} of gravitational radiation by perturbatively studying a subset of the equations of linearised gravity around Schwarzschild and propagating the asymptotics near past infinity towards $\Scrip$. 
The result of this preliminary analysis was that peeling also fails near future null infinity---the rigorous and complete analysis of this propagation problem is the topic of the present paper.

Further influential heuristics against smooth conformal compactification were put forth by Christodoulou in 2002 \cite{Chr02}. To this date, his argument remains the only argument treating the full non-linear Einstein equations without symmetry: Christodoulou suggests that any solution to \eqref{EVE} arising from suitable initial data must necessarily violate peeling near $\Scrim$ provided that there is no incoming radiation from $\Scrip$ and provided that the limit of the ingoing shear along $\Scrip$ (a.k.a.~the News function) decays as predicted by Einstein's quadrupole formula for a system of $N$ infalling masses following approximately hyperbolic Keplerian orbits in the infinite past. 

Note that this counterargument against previous notions of isolated systems can be interpreted to be itself based upon a different prescription of modelling isolated systems, the assumption being that isolated systems can be modelled by spacetimes arising from the initial data assumed in \cite{CK93} (cf.~the definition of C--K compatible initial data in Remark 1.1 of \cite{I}). 
In particular, these spacetimes are assumed to have certain decay towards spatial infinity $i^0$.

\begin{figure}[htpb]
\floatbox[{\capbeside\thisfloatsetup{capbesideposition={right,bottom},capbesidewidth=7cm}}]{figure}[\FBwidth]
{\caption{Schematic depiction of the generation problem (I), the propagation problem (II) and the asymptotic problem (III). If problems (I) and (II) are studied in the \textit{infinite past}, they determine the asymptotic behaviour (III) near $\Scrip$.\\
 In order to study them at \textit{finite} time, one instead needs to make an \textit{ad hoc} assumption about this asymptotic behaviour near $\Scrip$.}\label{fig:GPA}}
{\includegraphics[width=100pt]{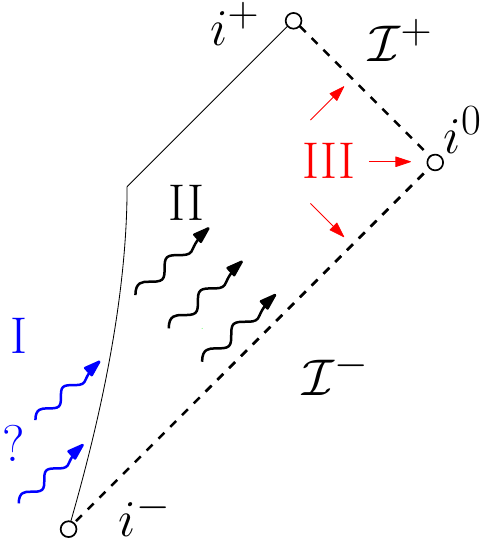}}
\end{figure}

The mathematically rigorous study of the case against smooth null infinity was then initiated in a series of papers of which the present one is the fourth. 
While the previous approaches towards modelling isolated systems either imposed certain asymptotic behaviour on gravitational radiation at some finite time, be it towards $\Scrip$ (peeling, Bondi coordinates...) or towards $i^0$ (as in \cite{CK93}), or by setting it to 0 as in the Bondi bomb spacetimes, the approach towards modelling isolated systems taken in this series is to let them arise as solutions to \textit{scattering problems in the infinite past}.
One of the current goals of this series of papers is the following:
We ultimately want to set up a scattering problem for the non-linear Einstein vacuum equations where the scattering data are informed by the Post-Newtonian prediction (the \textit{generation} problem\footnote{\label{fottynote}A longer-term goal that we won't address here is the development of a rigorous understanding of the \textit{generation problem}!}) for a system of $N$ masses coming in from the infinite past; and we then want to solve this scattering problem (i.e.~understand the \textit{propagation} problem) in order to understand the asymptotic structure of gravitational radiation near past \textit{and} future null infinity. See Fig.~\ref{fig:GPA}. 

As a precursor to this problem, we will first study the corresponding problem for the equations of linearised gravity around Schwarzschild.
The purpose of the current paper is to give an overview over this problem, its full and lengthy mathematical treatment being contained in follow-up work. 
We hope that we will be able to convey to the reader many of the main ideas (as the mechanism behind the failure of peeling is remarkably simple to describe) and results, and to also elucidate various points that we only touched upon in the exposition above. 

Since the series was initiated by an overview of Christodoulou's argument against peeling, we also reflect upon this argument, see \S\ref{sec:conclusion}.

\subsection{But why does this matter?}\label{sec:intro:why}
Before we move on to the main body of the paper, we want to put some effort into outlining a few reasons why the issue of null infinity being smooth or irregular is relevant.

Firstly, there is the epistemological level: If we have a widely-used and accepted concept that aims to model isolated systems, such as that of asymptotic simplicity, we should critically examine its justifications and shortcomings, and understand what kind of physics can, or can not, be captured by the concept. 
To be quite concrete, we will see in the main body of the paper that a spacetime can, somewhat ironically, essentially only be expected to have a smooth future null infinity if there is \textit{no outgoing radiation} (or a fine-tuned mix of outgoing and incoming radiation) in the infinite past; any radiation that reaches spatial infinity will generically lead to an irregular future null infinity.
Thus, operating under the assumption of a smooth null infinity excludes any physical scenario where there is radiation in the infinite past and effectively only leaves us with various types of Bondi bomb scenarios.\footnote{This entire discussion of course take place under the pretence that $\Lambda$ can reasonably be set to 0 and that the astrophysical process under consideration starts in the infinite past. The question of how to formalise the limiting process where $\Lambda$ is sent to 0 is yet a completely different story, and we do not dare to touch upon it here.}

Secondly, there is the physical level: Any physicist used to making approximations wherever possible should now feel inclined to point out that, from a measurement perspective, the violent outburst of gravitational radiation during, say, the merger of two black holes, should completely dominate the radiation that the black holes emitted in the ``infinite'' past (when they were ``infinitely far away from each other''), effectively also turning it into a Bondi bomb scenario.
However, as was observed in \cite{GK,II} (see also~\cite{Kroon01}), such intuitive arguments must necessarily become incorrect eventually: 
No matter the strength of any outbursts of gravitational radiation at finite time, if one measures for sufficiently long times, the late-time tails of gravitational radiation will be dominated by the effects of the irregularity of $\mathcal I^+$ (which in turn arises from the radiation in the infinite past).
This is because the late-time decay rates associated to radiation coming from the infinite past and radiation coming from some finite-time explosion are different, and the decay rate of the latter is faster. (See also~\S\ref{sec:late}.) 
Conversely, quantitatively studying the effects of gravitational radiation in the infinite past is exactly what allows one to control the error made by setting this very radiation to vanish.
The methods and results sketched in this paper can thus also be seen as tools  that can be used to try and justify the assumption of past-stationarity.

Thirdly, there is the mathematical level. 
The largest part of mathematical problems, e.g. concerning the stability or instability of explicit solutions such as the Minkowski spacetime, singular/regular behaviour in the interior of black holes, or the study of precise asymptotics of gravitational waves, are studied from the perspective of the Cauchy problem: Initial data are posed on an asymptotically flat (terminating at $i^0$) or an asymptotically null (terminating at $\Scrip$) hypersurface. 
One is thus forced to make an assumption on the asymptotic decay behaviour of the initial data set. Since there is no \textit{a priori} way to understand the asymptotic decay of an initial data set, any choice for this decay might thus be thought of as an \textit{ad hoc} choice of modelling a class of isolated systems.
Conversely, the present paper's approach towards modelling isolated systems taken here dynamically produces certain decay behaviour towards $i^0$ or towards $\mathcal I^+$, and will therefore affect all of the mathematical problems listed above.

For instance, the well-known and much studied Price's law (see e.g.~\cite{Price72,Leaver86,GPP94,AAG18a, AAG18b,AAG21,MZ21,Hintz22}) governing the late-time behaviour of gravitational radiation has mostly been studied under the assumption of compactly supported initial data (see~Fig.~\ref{fig:BB}). It has also been studied under the slightly weaker assumption of initial data satisfying peeling and recently, in \cite{II} and \cite{GK}, under assumptions that violate peeling, and it was shown that this crucially changes the late-time behaviour, cf.~our comment two paragraphs above. 
But since the precise late-time behaviour of gravitational radiation in the exterior of black holes affects the regularity in the interior of black holes, this, too, is influenced by the regularity of $\Scrip$ (see e.g.~\cite{Dafermos05interior,LukOh16,Sbierski22,MZ22}).

The situation is similar for stability studies of explicit solutions to \eqref{EVE}:
Let's recall the monumental work on the stability of Minkowski spacetime \cite{CK93}. In this work, Christodoulou and Klainerman consider perturbations of the Minkowski initial data that satisfy inter alia $g_{ij}=(1+\tfrac{2M}{r})\delta_{ij}+o(r^{-3/2})$ ($g$ denoting the metric on the initial data surface $\{t=0\}$)), and show that the corresponding solutions disperse and asymptotically approach the Minkowski spacetime with quantitative rates. 
Since these quantitative rates were slower than predicted by~\eqref{eq:intro:peeling}, 
the results of \cite{CK93} were, in particular, incorrectly interpreted by some to disprove peeling, even though \cite{CK93} only showed that there are certainly classes of spacetimes with decay behaviour towards $\Scrip$ weaker than peeling---the question whether peeling is ``true'' or not can only be answered by taking into account physical arguments. 
And indeed, as we will see in the present paper, the initial decay $g_{ij}=(1+\tfrac{2M}{r})\delta_{ij}+o(r^{-3/2})$ is itself too strong to model certain physically relevant perturbations---the present work constructs spacetimes which, restricted to $t=0$, asymptotically decay like
\begin{equation}
g_{ij}=(1+\tfrac{2M}{r})\delta_{ij}+\O_{\ell\leq 1}(r^{-2})+\O_{\ell\geq 2}(r^{-1}).
\end{equation}
Out of all the various proofs of stability of the Minkowski spacetime floating about in the literature (e.g.~\cite{CK93,KlNi03,LR10,Keir18,HV20,Hintz23} and references therein), only Bieri's class of perturbations \cite{Bieri} and those of \cite{IonescuPausader22,Shen23} are large enough to allow for such slow decay---all other proofs assume in particular that $g=(1+\tfrac{2M}{r})\delta+o(r^{-1})$!

Somewhat surprisingly, even though we must thus conclude that most stability works starting from a Cauchy hypersurface make assumptions too strong to model a physically relevant class of perturbations, we will see that these very perturbations exhibit sufficient decay towards $\Scrip$ so as to still be compatible with most stability works that start from an asymptotically null initial data hypersurface \cite{DHR16, KlSz21, DHRT21}! See \S\ref{sec:conclusion}.  

\subsection{Structure of the remainder of the paper}
In \S\ref{sec:phys}, we give an outline of the Post-Newtonian analysis of the \textit{generation problem} for a system of $N$ infalling masses coming from the infinite past.

In \S\ref{sec:math}, we describe how this Post-Newtonian analysis informs the mathematical setup of a scattering problem for the linearised Einstein vacuum equations around Schwarzschild, i.e.~the mathematical setup for the \textit{propagation problem}.

In \S\ref{sec:scat} and \S\ref{sec:asy1}, we describe how this scattering problem is solved, and we study the asymptotic properties of the Weyl curvature tensor of the scattering solution. In particular, we give a quick and intuitive sketch of why peeling fails in the way it does.

In \S\ref{sec:asy2}, starting from the asymptotics of the Weyl tensor, we sketch how to derive the asymptotics of the remainder of the system and comment on various interesting properties that the resulting solutions have.
We briefly discuss consequences of our results on late-time asymptotics in \S\ref{sec:late} and conclude in \S\ref{sec:conclusion}.

\section{The physical setup for the generation problem}\label{sec:phys}
We here sketch the physical setup that will inform our mathematical construction. 
This section does not aim or claim to be rigorous, instead, we want to give a ``quick and dirty'' justification for the assumptions we will make for our mathematical scattering setup in \S\ref{sec:math}.
Some conceptual difficulties are mentioned in \S\ref{sec:phys:prob}.
\subsection[The physical picture that we want to describe]{The physical picture that we want to describe\dots}\label{sec:phys:pic}
\dots is that of a system of two infalling masses (with negligible internal structure) whose trajectories approach hyperbolic (or parabolic) Keplerian orbits in the infinite past. 
More generally, the contents of this section apply to systems of $N$ infalling masses from the infinite past whose relative velocities approach constant, nonrelativistic values.
In addition, we prohibit incoming radiation from $\Scrim$.
By virtue of these masses moving slowly and the separation between them becoming large, we expect perturbations around the Newtonian theory to provide us with a good prediction on the gravitational radiation generated by this system.

\begin{figure}[htpb]
\floatbox[{\capbeside\thisfloatsetup{capbesideposition={right,bottom},capbesidewidth=5cm}}]{figure}[\FBwidth]
{\caption{We apply perturbative methods to study the generation of gravitational waves up until some null cone $\Cinn$. The region beyond $\Cinn$ will later be studied rigorously by taking the behaviour along $\Cinn$ as given.}\label{fig:PS}}
{\includegraphics[width=220pt]{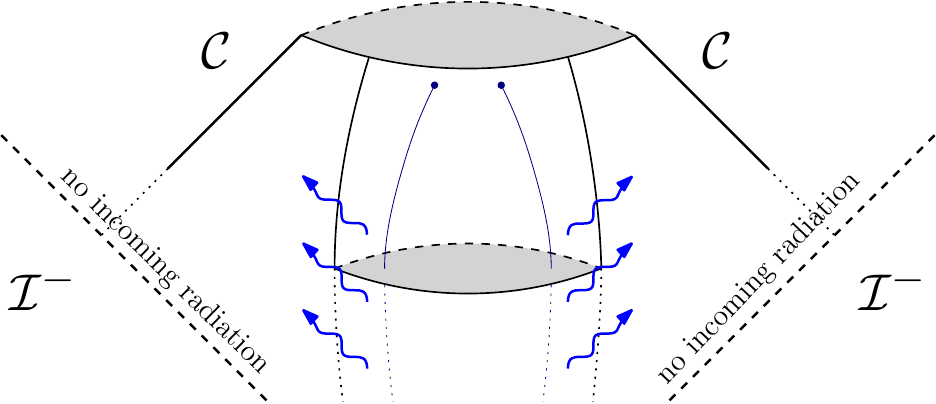}}
\end{figure}

The plan is now as follows:
We enclose the system by an ingoing null cone $\Cinn$ from $\Scrim$, truncated at some finite retarded time since we only care about the behaviour of the system in the distant past. Up until this null cone (we denote this region \textit{the intermediate region}), we assume the validity of a suitable Post-Newtonian framework in order to understand the \textit{generation problem}, in particular, we use this framework to get a prediction on the behaviour of gravitational radiation along $\Cinn$ towards $\Scrim$. See Fig.~\ref{fig:PS}.

\subsection{The multipolar Post-Minkowskian expansion and the Post-Newtonian prediction}
The generation of gravitational waves by Newtonian or Post-Newtonian sources has been studied in early references, see, for instance, \cite{WagonerEpstein75,WagonerWill76} (and \cite{Blanchet14} for a more recent review). 
The physical picture of two weakly interacting masses in the infinite past as described in \S\ref{sec:phys:pic} was first studied to leading order in the Post-Newtonian expansion parameter $v/c$  by Walker and Will \cite{WalkerWill79I,WalkerWill79}, $v$ denoting the characteristic speed of the system and $c$ the speed of light. This expansion corresponds to an expansion that only takes into account the quadrupole moment of gravitational radiation, higher multipole moments being of higher order in $v/c$.
The relevant results can be read off from eqns.~(61), (62) in \cite{WalkerWill79}.
In particular (the reason why we single out the following results will be made clear in \S\ref{sec:math}), the following asymptotic behaviour was found along the incoming Minkowskian null cones:
$\Psi_0$ was found to decay like $r^{-3}$ towards $\Scrim$, $\Psi_4$ was found to decay like~$r^{-4}$ (violating peeling~\eqref{eq:intro:peelinpast}), and the ingoing shear along the null cones, $\lambda$, was found to decay like $r^{-2}$.
In each case, the coefficient of the relevant decay rate is related to (time derivatives of) the quadrupole moment of the system, more precise expressions will be given below.

In order to obtain a more complete picture, we also want to understand the asymptotic behaviour of higher multipoles; for this, we consult Thorne's work on multipole expansions in GR \cite{Thorne80} (see also the appendix of \cite{SeriesVI} or Pirani's lecture notes \cite{Pirani64}).

Let's give a quick summary of the computations in \cite{Thorne80}: We will temporarily work in Cartesian Minkowskian coordinates $(x^0=t,x^1,x^2,x^3)$, and we will take Latin indices to only range from 1 to 3.
First, Thorne considers linearised perturbations $g_{\mu\nu}^1=g_{\mu\nu}-\eta_{\mu\nu}$ around the Minkowski metric $\eta_{\mu\nu}$. 
By imposing de Donder gauge, the linearised Einstein vacuum equations then reduce to Minkowskian wave equations for the Cartesian components of the trace-reversed metric perturbation $g^1_{\mu\nu}-\tfrac12 \eta_{\mu\nu} \eta^{\alpha\beta} g^1_{\alpha\beta}$, and the general \textit{outgoing wave} solution for $g^1$, assuming the coordinates to be mass-centered, is then given in equations~(8.13) of \cite{Thorne80} in terms of an expansion into multipoles $\mathscr{I}_{A_{\ell}}$ and $\mathscr{S}_{A_{\ell}}$, the radiation field corresponding to this solution being given by eq.~(4.8) of \cite{Thorne80}. 

Let now $(t,r,\theta,\varphi)$ denote the usual spherical polar coordinates, introduce the double null coordinates $u=t-r$, $v=t+r$, define the null tetrad $l=\partial_t+\partial_r=2\partial_v$, $n=\partial_t-\partial_r=2\partial_u$, $m_{\pm2}=\tfrac{1}{\sqrt{2}r}(\partial_\theta\pm i\tfrac{\partial_\varphi}{\sin\theta})$, and then define the Newman--Penrose scalars
\begin{align}\label{eq:phys:defPsi0Psi4}
\Psi_0=W^1_{\mu\nu\rho\sigma}l^\mu m_2^\nu l^\rho m_2^\sigma,&&\Psi_4=W^1_{\mu\nu\rho\sigma}n^\mu {m}_{-2}^\nu n^\rho {m}_{-2}^\sigma,
\end{align}
where $W^1$ denotes the linearised Weyl tensor of $g$. 
From the form of the metric (8.13), we then compute (this generalises~\cite{JanisNewman65,Lamb66}),
for $s=\pm2$
\begin{nalign}\label{eq:phys:Psi0Psi4spin}
\Psi_{2-s}=-2^{s}\sum_{\ell=2}^{\infty}\sum_{m=-\ell}^{\ell}\frac{(\ell+s)!}{(\ell-2)!}r^{\ell-|s|}\partial_u^{\ell-s}\left(\frac{I_{\ell m}(u)T_{ij}^{\mathrm{E}2,\ell m} -\frac{s}{|s|} S_{\ell m}(u)T_{ij}^{\mathrm{B}2,\ell m}}{r^{1+\ell+s}}\right)m_s^i m_s^j,
\end{nalign}
where the $I_{\ell m}$ and $S_{\ell m}$ are related to $\mathscr{I}_{A_\ell}$ and $\mathscr{S}_{A_{\ell}}$ via Thorne's (4.7) and $T_{ij}^{\mathrm{E}2,\ell m}$ and $T_{ij}^{\mathrm{B}2,\ell m}$ denote the electric and magnetic transverse traceless tensor  spherical harmonics defined in Thorne's (2.30). (See Thorne's eq.~(2.38) for the relation to spin-weighted harmonics.)

By now considering slow-motion, weak internal gravity sources in the ``near zone'' of spacetime, by using the outgoing Minkowskian Green's function and by then matching it to the vacuum expression for the metric (8.13), it is derived that, to leading order in $v/c$, the moments $I_{\ell m}$ and $S_{\ell m}$ are given by the Newtonian mass and current multipole moments (eq. (5.27) of \cite{Thorne80}):
\begin{nalign}\label{eq:phys:Newtonmoments}
I_{\ell m}&=\tfrac{16\pi}{(2\ell+1)!!}\sqrt{\tfrac{(\ell+1)(\ell+2)}{2(\ell-1)\ell}} \int \rho\cdot r^{\ell} Y^{\ell m\ast}\dd ^3 x,\\
S_{\ell m}&=\tfrac{32\pi}{(2\ell+1)!!}\sqrt{\tfrac{(\ell+2)}{2(\ell-1)}} \int (\epsilon_{jpq}x_p\cdot\rho v_q)\cdot r^{\ell-1}Y_j^{\mathrm{B},\ell m\ast}\dd ^3x.
\end{nalign}
In the formulae above, $\epsilon$ denotes the Levi-Civita symbol, $\rho$ and $\rho v_j$ denote the Newtonian mass and momentum density, and $Y^{\mathrm{B},\ell m}$ is the magnetic vector spherical harmonic (defined in Thorne's (2.18)).

With these computations at hand, we can now extend the results of \cite{WalkerWill79} for the setting of two pointlike particles moving along approximately hyperbolic Keplerian orbits beyond the mass quadrupole approximation.
 For such orbits, the leading order behaviour of the relative position vector $x_{\mathrm{rel}}(t)$ of the particles is given by $x_{\mathrm{rel}}^i(t)=v^i t+a^i \log |t|+b^i+\dots$ as $t\to-\infty$ (see \cite{Eder83} for a similar statement for $N$ particles).
Here, the coefficient $v$ corresponds to the asymptotic relative velocity of the orbit, and $|a|\sim (m_1+m_2)/v^2$ by Newton's equations of motion.

If we insert this asymptotic behaviour for $x_{\mathrm{rel}}(t)$ into the multipole moments \eqref{eq:phys:Newtonmoments}, we find that
$
I_{\ell m}(u)=I_{\ell m}^0u^\ell+I_{\ell m}^1u^{\ell-1}\log |u|+\dots,
$
and $
S_{\ell m}(u)=S_{\ell m}^0u^\ell +S_{\ell m}^1 u^{\ell-1}\log |u|+\dots,
$
where $|I_{\ell m}^0|$ is of magnitude $(v/c)^\ell$, and where $|S_{\ell m}^0|\sim (v/c)^{\ell+1}$. 
We conclude from this and \eqref{eq:phys:Psi0Psi4spin} that, as we approach $\Scrim$, all $\ell$-modes $\Psi_{2\pm2,\ell}\sim r^{\ell-2} \partial_{u}^{\ell\pm2}(\tfrac{u^{\ell}+u^{\ell-1}\log|u|}{r^{\ell+1\mp2}})$ of the Newman--Penrose scalars $\Psi_{2\pm2}$ decay like:
\begin{align}\label{eq:phys:theprediction}
\Psi_{0,\ell}=\frac{\Psi_{0,\ell}^0}{ r^{3}}+\frac{\Psi_{0,\ell}^1\log r}{r^{4}}+\frac{\Psi_{0,\ell}^2}{r^4}+\dots,
&&\Psi_{4,\ell}=\frac{\Psi_{4,\ell}^0}{ r^{4}}+\dots,
\end{align}
where $\Psi_{0,\ell}^0$ is computed from $I_{\ell m}^0$ (and $S_{\ell m}^0$), $\Psi_{0,\ell}^1$ is computed from $I_{\ell m}^1$ (and $S_{\ell m}^0$) and, up to complex conjugation and $\ell$-dependent multiple, $\Psi_{0,\ell}^1\sim \Psi_{4,\ell}^{0}$. In the above, we used that $|u|\sim r$ near $\Scrim$.
It is also straight-forward to compute that all higher $\ell$-modes of the ingoing shear decay as for $\ell=2$ (cf.~eq.~(63) of~\cite{WalkerWill79}), i.e.~$r^2\lambda_\ell$ attains a finite, nonzero limit at $\Scrim$. 

A similar story can be told for parabolic orbits, where $x_{\mathrm{rel}}(t)\sim t^{\frac{2}{3}}$. 
The behaviour of the $\ell=2$ modes of $\Psi_0$ and $\Psi_4$ near $\Scrim$ is then given by $r^{-\frac{11}{3}}$. 
In contrast to hyperbolic orbits, since parabolic orbits do not grow linearly in time, higher $\ell$-modes will now decay faster towards $\Scrim$: $\Psi_{2\pm2,\ell}\lesssim r^{-3-\ell+\frac{2\ell}{3}}$.
\subsection{Comments and conceptual problems}\label{sec:phys:prob}
We conclude with some comments. 
First, notice that \eqref{eq:phys:Psi0Psi4spin} implies that $\Psi_0=\O(r^{-5})$ near $\Scrip$, independently of the structure of the multipole moments. 
Historically, this fact was used as justification for the peeling property \eqref{eq:intro:peeling}, e.g.~in \cite{NP62Approach}. Similarly, the fact that outgoing wave solutions to the linearised field equations around Minkowski have an expansion in powers of $1/r$ (see Thorne's (8.13)) motivated the setup of Bondi coordinates.

What we will see in the present note, however, is that Thorne's expression for the metric perturbation (8.13) breaks down at late advanced times, i.e.~near $\Scrip$, due to the effects of the Schwarzschild term $M/r$. In fact, taking into account this Schwarzschild term will also lead to corrections in Einstein's classical quadrupole formula along $\Scrip$ ((4.18) of \cite{Thorne80}).
Of course, one might then wonder whether it is already too much to assume the validity of the framework all the way until the null cone $\Cinn$, rather than, say, just assuming it to hold up until some timelike cylinder $\mathcal T$ along which $|t|\sim r$ as in \cite{Damour86}, see Fig.~\ref{fig:PS}.
Let us here just mention that the propagation of radiation from such a timelike cylinder to the null cone $\Cinn$ was studied at the level of scalar waves in \cite{I,III} on Schwarzschild: the result of these studies is that the leading order decay towards $u\to-\infty$ is the same along the cylinder and along $\Cinn$. 

Secondly, the system that we consider, growing linearly in time, has the feature that all higher-order corrections within the Post-Newtonian expansion (see \S V.D of \cite{Thorne80}), even though they are of smaller magnitude in $v/c$, feature the same decay towards~$\mathcal I^-$, in just the same way as all electric angular modes of $\Psi_0$ or $\Psi_4$ feature the same decay towards $\mathcal I^-$. 
Of course, one could interpret the approximate statements \eqref{eq:phys:Newtonmoments} as the leading order result for metric perturbations on fixed angular frequency. 
If we really wanted to do consistent perturbation theory, however, then writing down the leading-order expressions \eqref{eq:phys:Newtonmoments} for the $\ell$-th electric multipole moment means that we should also consider higher-order Post-Newtonian expansions of all lower electric multipoles (i.e.~expand the $(\ell-n)$th multipole up to $n$ orders) along with higher-order Post-Minkowskian expansions around $\eta$ etc.

Lastly, while we do not expect that the effects of higher-order perturbations will change the decay rates of \eqref{eq:phys:theprediction} (they will certainly change the coefficients!), it would certainly be an interesting problem to understand this in more detail, cf.~footnote~\ref{fottynote}.

\newcommand{\LG}{$\textbf{\textit{(LGS)}}$}
\section{The mathematical setup for the propagation problem}\label{sec:math}
In the previous section, we gave a rough sketch of the \textit{generation problem} of gravitational radiation  in a Post-Newtonian framework.
We now describe how we convert the information from \S\ref{sec:phys}, which was perturbatively obtained for linearised gravity around Minkowski, into a mathematically formulated scattering setup for linearised gravity around Schwarzschild. 

Studying linearised gravity around Schwarzschild will on the one hand provide us with a clear understanding of the failure of peeling due to the presence of mass near spatial infinity.
On the other hand, we will set up the problem in such a way as to serve as a strong foundation for the study of the actual, nonlinear Einstein vacuum equations.

\subsection{The system of linearised gravity around Schwarzschild \texorpdfstring{\LG}{(LGS)}}
Consider the Schwarzschild metric in the familiar form $g_M=-(1-\tfrac{2M}{r})\dd t^2+(1-\tfrac{2M}{r})^{-1}\dd r^2+r^2(\dd \theta^2+\sin^2\theta\dd \varphi^2)$.
 We introduce the double null coordinates $u=(t-r^*)/2$, $v=(t+r^*)/2$, where $r^*=r+2M\log (r/2M)$, so that the metric takes the form
\begin{equation}
g_{M}=-4\Omega^2\dd u\dd v+\slashed{g}_{AB}\dd \theta^A\dd \theta^B.
\end{equation}
Here, $\Omega^2=1-\tfrac{2M}{r}$, and $\slashed{g}$ denotes the standard metric on $\mathbb S_r^2$, with $(\theta^1, \theta^2)=(\theta,\varphi)$. 
(In the following, capital Latin indices will always refer to indices on the sphere and range from 1 to~2.)
We will restrict our attention to the very exterior of spacetime, that is to say, we will work on the manifold $\mathcal M_M=(-\infty,-1]_u\times [1,\infty)_v\times \mathbb S^2$ (see Fig.~\ref{fig:MS}); and we will work with the ON frame
\begin{equation}
\left(e_1=\tfrac{1}{r}\partial_\theta,e_2=\tfrac{1}{r\sin\theta}\partial_{\varphi},e_3=\tfrac{1}{\Omega}\partial_u,e_4=\tfrac{1}{\Omega}\partial_v\right).
\end{equation}

The (double null) system of linearised gravity around Schwarzschild is obtained by considering a general 1-parameter family of Einstein metrics $\bm{g}(\varepsilon)$ on $\mathcal M_M$ such that $\bm{g}(0)=g_M$:
\begin{equation}\label{eq:math:doublenull}
\bm{g}(\epsilon)=-4\bm{\Omega}^2(\varepsilon) \dd u\dd v+\bm{\slashed{g}}_{AB}(\varepsilon)(\dd \theta^A- \bm{b}^A(\epsilon)\dd v)(\dd \theta^B- \bm{b}^B(\epsilon)\dd v);
\end{equation}
 and by then linearising the arising Einstein and Bianchi equations in $\varepsilon$ (i.e.~by writing $\bm{\Omega}=\sqrt{1-\tfrac{2M}{r}}+\varepsilon \Om$ etc.~and only keeping terms of order $\varepsilon$). See \cite{DHR16} for details.
 
The result is a coupled system of 10 hyperbolic equations for the linearised curvature coefficients---the linearised Bianchi equations---together with a system of around 30 transport and elliptic equations for the linearised metric and connection coefficients. We will refer to this system as \LG.

Examples for linearised metric components are the linearised lapse $\Om$ and the linearised metric on the spheres $\gs$, which is split up into its trace and its tracefree part, $\gs=\gsh+\tfrac{1}{2}\trg\cdot \slashed{g}$. 
 
  Examples for linearised connection coefficients are the linearisation of the ingoing and outgoing null expansions, denoted $\trxb$ and $\trx$, respectively, and of the ingoing and outgoing null shears: $\xhb$ and $\xh$. 
The work~\cite{DHR16} employs the Christodoulou--Klainerman formalism. For those familiar with the very closely related Newman--Penrose formalism, the ingoing and outgoing null expansions (or shears) are called $\mu$ and $\rho$ (or $\lambda$ and $\sigma$), respectively.

Examples for linearised curvature coefficients are the linearised Gauss curvature $\K$ on the spheres, as well as the extremal curvature components $\alpas[-2]$ and $\alpas[+2]$:
\begin{align}
\alpas[-2](e_A,e_B)=\Omega^2\overone{W}(e_3,\e_A,e_3,e_B),&& \alpas[+2](e_A,e_B)=\Omega^{-2}\overone{W}(e_4,e_A,e_4,e_B).
\end{align}
These are the real-valued symmetric tracefree two-tensor analogues to $\overone{\Psi}_4$ and $\overone{\Psi}_0$: 
Identifying $l=\Omega e_4$, $n=\Omega^{-1} e_3$, $\sqrt{2}m_{\pm2}=e_1\pm ie_2$, we have $\overone{\Psi}_{2\mp2}=\alpas[\pm2](m_{\pm2},m_{\pm2})$.

Examples for equations of \LG~are given by (writing $\xh_{AB}=\xh(e_A,e_B)$ etc.)
\begin{align}\label{eq:math:exampleequations}
\pu(\gsh_{AB})=2\Omega \xhb_{AB},\quad \pu(\Omega^{-1}r^2\xhb_{AB})=-\Omega^{-2}r^2\alpas[-2]_{AB},\quad \pv(\Omega^{-1}r^2\xh_{AB})=-\Omega^{2}r^2\alpas[2]_{AB}.
\end{align}

\subsection{The pure gauge solutions and the linearised Kerr solutions of \texorpdfstring{\LG}{(LGS)}}\label{sec:math:gauge}
Central to the understanding of \LG~is the existence of two classes of explicit solutions to it.
The first class of these consists of the \textbf{linearised Kerr solutions}: These are produced by linearising a nearby Kerr metric in double null coordinates around Schwarzschild, or by linearising a Schwarzschlid solution with nearby mass around the original Schwarzschild.

The other class of solutions is given by the \textbf{pure gauge solutions}. These are generated by linearising nonlinear coordinate transformations of the double null coordinates $(u,v,\theta^1,\theta^2)$ that preserve the double null form of the metric \eqref{eq:math:doublenull}. 

It turns out that the linearised Kerr solutions are entirely supported on spherical harmonics with $\ell<2$. 
Conversely, any solution supported on $\ell\leq1$ is given by a linear combination of such linearised Kerr and pure gauge solutions. 
In other words, the $\ell\leq1$-part of \LG~doesn't carry any dynamics. 
In what follows, we will therefore exclusively discuss solutions supported on $\ell\geq 2$.

In view of the existence of these solutions, it is very helpful to extract quantities from the system \LG~that are invariant under addition of pure gauge or Kerr solutions. 
Examples for such \textit{gauge invariant} quantities are $\alpas[+2]$ and $\alpas[-2]$ (i.e.~$\overone{\Psi}_0$ and $\overone{\Psi}_4$). 
In fact, any solution that has both $\alpas[-2]$ and $\alpas[+2]$ (or $\overone{\Psi}_4$ and $\overone{\Psi}_0$) vanishing is given by a combination of linearised Kerr and pure gauge solutions.
See \cite{WaldTeuk,DHR16} for details.

\subsection{Scattering theory for \texorpdfstring{\LG}{(LGS)}: The seed scattering data}\label{sec:math:seed}
The quantitative stability of \LG~was first shown in \cite{DHR16} (and provided the central ingredient to the more recent nonlinear stability of Schwarzschild proof \cite{DHRT21}), and a global scattering theory to \LG~was written down in~\cite{Masaood22b}. 

\begin{figure}[htpb]
\floatbox[{\capbeside\thisfloatsetup{capbesideposition={right,bottom},capbesidewidth=6cm}}]{figure}[\FBwidth]
{\caption{We rigorously study the propagation problem to the future of the null cone $\Cinn$ by formulating a scattering problem for \LG~in $D^+(\Cinn\cup\Scrim)$. We always think of the scattering data along $\Cinn$ as capturing the radiation of some physical system, cf.~Fig.~\ref{fig:PS}.}\label{fig:MS}}
{\includegraphics[width=120pt]{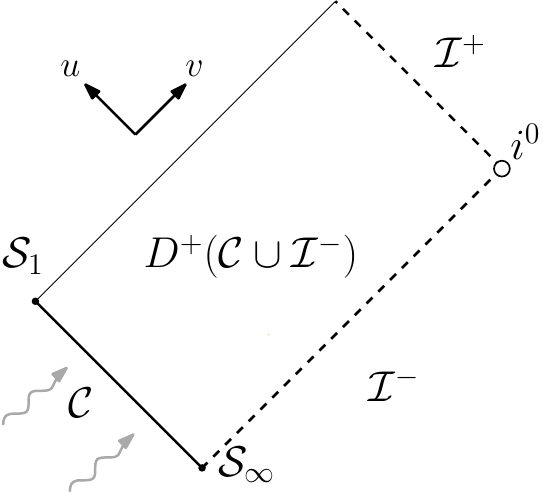}}
\end{figure}

In the present paper, we will be concerned with the \textit{semi-global scattering theory} for \LG: 
Given appropriate data on the truncated ingoing null cone $\Cinn=\mathcal M_{M}\cap\{v=1\}$ (whose future end sphere we will denote as $\Sone$ and whose past limiting sphere we will denote as $\Sinfty$, see Fig.~\ref{fig:MS}) coming from $\Scrim$ together with data along $\Scrim$ to the future of this cone, we want to construct the unique solution to \LG ~``restricting'' to these data in the limit.

We now describe what scattering data consist of, and which parts of these scattering data carry physical information. 
As the $\ell<2$-part \LG~is non-dynamical, we will only present the discussion for data and solutions supported on $\ell\geq 2$.

\begin{definition}\label{def:seed}
An \emph{$\ell\geq2$ seed scattering data set} consists of the following prescribed quantities:
Along~$\Cinn$, prescribe $\xhb$ and $\pu\Om$. 
Along $\Scrim$ (to the future of $\Sinfty$), prescribe $\gsh$, $\Om$ and the radiation field of the outgoing null shear $r\cdot \xh$ (a.k.a.~the News tensor).
Finally, prescribe on $\Sone$ the values of $\pv\alpas[-2]$ as well as the ingoing null expansion $\trxb$, and prescribe on $\Sinfty$ the weighted outgoing null expansion $r\cdot \trx$ together with the Gauss curvature $r^2\K$.
\end{definition}

From these seed scattering data, one can construct all other quantities (e.g.~$\alpas[-2]$ or $\alpas[+2]$) along~$\Cinn$ and $\Scrim$, and then construct the unique solution to \LG~in the entire domain of dependence $D^+(\Cinn\cup\Scrim)$ that restricts to these data (in the limit).
A sketch of this is given in~\S\ref{sec:scat}.
\subsection{Bondi normalisation of the seed scattering data}\label{sec:math:Bondi}
We now discuss which part of the seed data carry physical meaning. 
For instance, while $r\xh$ along $\Scrim$ carries radiative physical information, $\gsh$ along $\Scrim$ (\textit{without} the extra $r$-weight!) only describes information related to the choice of angular coordinates at infinity that can always be made to vanish by a gauge transformation.
More generally, in \cite{V}, it is shown that, upon adding pure gauge solutions, any seed scattering data set can be Bondi normalised at $\Scrim$; this means:
\begin{itemize}
\item $\gsh$ and $\Om$ along $\Scrim$ can be set to zero, and
\item $r^2\K$ and $r\trx$ on $\Sinfty$ can be set to zero.
\end{itemize}
Together, the two bullet points imply that $r^2\K$ and $r\trx$ vanish along all of $\Scrim$, meaning that the spheres at $\Scrim$ are the standard round spheres. 
This normalisation captures the spirit of the original Bondi coordinates, except that it does not make any assumptions on $1/r$-expansions.
Henceforth, we will always assume our seed data to be Bondi normalised.

The Bondi normalisation as described above still leaves us with some remaining gauge freedom which, for example, contains the group of BMS transformations on $\Scrim$, see \cite{Masaood22b} for details. 
The remaining gauge freedom can also be used to set $\pu\Om$ along $\Cinn$ and $\trxb|_{\Sone}$ to~0. 
Doing this, we see that the physical content of the seed scattering data is carried by $r\xh|_{\Scrim}$, $\xhb|_{\Cinn}$ as well as $\pv\alpas[-2]|_{\Sone}$.

\subsection{Seed scattering data describing the exterior of the \texorpdfstring{$N$}{N}-body problem}\label{sec:phys:physicaldata}
Having understood what constitutes scattering data for \LG, we now want to pose scattering data as motivated by the considerations in \S\ref{sec:phys}.
What we would like to do is to just take the statements \eqref{eq:phys:theprediction} and demand that our seed scattering data are such that the predicted asymptotics for $\alpas[-2] (=\overone{\Psi}_4)$ and $\alpas[+2](=\overone{\Psi}_0)$ near $\Scrim$  are satisfied.
But of course, the results of \S\ref{sec:phys} concern linearised gravity around Minkowski and, as such, do not translate directly into results for linearised gravity around Schwarzschild. 
Complications in relating the two include the logarithmic divergence of the null cones in Schwarzschild compared to Minkowski, metric perturbations around Schwarzschild being different objects from metric perturbations around Minkowski, as well as changes to the treatment of the near zone generating the relationship to the Newtonian theory.
As these complications are quite severe, we will not attempt to treat them, but instead just hope that all of these changes are subleading in terms of decay towards $\Scrim$.
On the basis of this hope, we formulate the following definition:
\begin{definition}\label{def:physicalseed}
A Bondi normalised seed scattering data set is said to describe \emph{the exterior of a system of $N$ infalling masses following approximately hyperbolic orbits} if the following conditions are satisfied for all $\ell\geq2$:
\begin{enumerate}[label=(\Roman*)]
\item $\alpas[-2]$ along $\Cinn$ satisfies $\alpas[-2]_{\ell}=\albdata_{\ell} r^{-4}+\O(r^{-4-\epsilon})$ for some $\albdata_{\ell}\neq 0$ independent of $u$, and the limit $\lim_{u\to-\infty}r^2\xhb_{\ell}$ along $\Cinn$, which is finite by the second of \eqref{eq:math:exampleequations}, is nonvanishing.\label{ListI}
\item The limit $\lim_{u\to-\infty}r^3\alpas[+2]_{\ell}=\aldata_{\ell}$ is finite and nonvanishing.\label{ListII}
\end{enumerate}
Finally, we say that such data also satisfy the \emph{no incoming radiation condition from $\Scrim$} if additionally the News tensor $r\xh$ vanishes along $\Scrim$.
\end{definition}
In the above definition, $\alpas[+2]_{\ell}$ denotes the projection of $\alpas[+2]$ onto the electric and magnetic tensor harmonics $T^{\mathrm{E2},\ell m}$ and $T^{\mathrm{B2},\ell m}$ introduced below \eqref{eq:phys:Psi0Psi4spin}. We will occasionally use an overline to denote magnetic conjugation:
Given $\aldata=\aldata^{\mathrm{E}2}+\aldata^{\mathrm{B}2}$, $\overline{\aldata}$ denotes $\overline{\aldata}=\aldata^{\mathrm{E}2}-\aldata^{\mathrm{B}2}$.

The assumptions \ref{ListI} and \ref{ListII} are directly motivated by eq.~\eqref{eq:phys:theprediction} and the corresponding result for~$\lambda_\ell$ below~\eqref{eq:phys:theprediction}, whereas the last condition of Def.~\ref{def:physicalseed} realises the physical requirement of excluding incoming radiation from $\Scrim$, it is the linearised version of demanding that the Bondi mass along~$\Scrim$ be constant.

It is, of course, possible to formulate the conditions of Def.~\ref{def:physicalseed} directly in terms of the seed data---see \cite{V} (in fact, for \ref{ListI}, the reader can see this from \eqref{eq:math:exampleequations}). 
\subsection{The main result}
We already formulate a rough version of one of the main results of \cite{V}:
\begin{theorem}\label{thmIV}
There exists a unique solution to \LG~attaining the scattering data of Def.~\ref{def:physicalseed}. This solution satisfies, throughout $D^+(\Cinn\cap\Scrim)$:
\begin{multline}\label{eq:thm1}
\alpas[+2]_\ell=\frac{1}{r^5}\sum_{n=0}^{\ell-2}f_{n,\ell}(u)\left(\frac{|u|}{r}\right)^n \\
+\frac{2M(-1)^{\ell+1}}{(\ell-1)(\ell+2)}\cdot \frac{\aldata_{\ell}}{r^4}+\frac{M(-1)^{\ell+1}(\ell-1)(\ell+2)}{2}\frac{\overline{\albdata}_{\ell}\log^2 r}{r^5}+ \O(Mr^{-5}\log r) 
\end{multline}
Here, $f_{n,\ell}(u)=C_{n,\ell} \aldata_{\ell}u^2+\O(|u|\log|u|)$ for some algebraic constants $C_{n,\ell}$.

Moreover, the solution satisfies, as $u\to-\infty$,
\begin{equation}\label{eq:thm2}
\lim_{u=\mathrm{const},v\to\infty}r\alpas[-2]_\ell=(-1)^{\ell+1}\cdot 2|u|^{-3}\left(\frac{\albdata_{\ell}}{\ell(\ell+1)}+\frac{2M\overline{\aldata}_{\ell} (\ell-2)!}{(\ell+2)!}\right)+\O(|u|^{-3-\epsilon}).
\end{equation}
\end{theorem}
The proof of this result will be sketched in \S\S\ref{sec:scat}--\ref{sec:asy1}, and further commentary on the result will be provided in the sections afterwards.
\section{Solving the scattering problem I: Constructing the unique solution}\label{sec:scat}
\newcommand{\Ph}{\overone{\Phi}}
In \S\ref{sec:math}, we outlined the mathematical setup of the scattering problem for \LG~and related this setup to the generation of gravitational waves discussed in \S\ref{sec:phys}. 
We will now sketch the general construction of scattering solutions to \LG. 
\subsection{The Teukolsky equations and the Regge--Wheeler equation}
The system \LG~may at first seem quite untractable, on the one hand owing to its size (it consists of around 40 equations for 19 unknowns), on the other hand due to the large space of pure gauge solutions. 
A key realisation in unlocking \LG, bypassing both difficulties, is that the gauge invariant quantities $\alpas[+2]$ and $\alpas[-2]$ each satisfy a decoupled wave equation known as the Teukolsky equation, which, in components ($\alpas_{AB}=\alpas(e_A,e_B)$), reads:
\begin{multline}\label{eq:Teuks}
\pu(r^{-2s}\Omega^{2s}\pv(r^{|s|+s+1}\alpas_{AB}))\\
=\frac{\Omega^{2s+2}}{r^{2s+2}}r^{|s|+s+1}((\lap+s)\alpas)_{AB}-\frac{2M\Omega^{2s+2}}{r^{2s+3}}(1+s)(1+2s)r^{|s|+s+1}\alpas_{AB},\quad\,s=\pm2.
\end{multline}
Here, $\lap$ denotes the Laplacian on $\Stwo$ acting on a two-tensor, its eigenvalues are $-\ell(\ell+1)+4$, $\ell\in\mathbb{N}_{\geq2}$. Henceforth, we will by $\pv\alpas$ or $\pu\alpas$ always mean the derivative of the components of $\alpas[+2]$ (i.e.~$\pv(\alpas(e_A,e_B))$), whereas, by $\lap\alpas$, we will mean the components of $\lap$ acting on a two-tensor (i.e.~$(\lap\alpas)(e_A,e_B)$).

Despite the large body of literature on linear wave equations on black hole backgrounds, a robust analysis of the Teukolsky equation for a long time seemed inaccessible due to the first order terms that appear when writing \eqref{eq:Teuks} in standard wave form ($\pu\pv\alpas=\dots$).
Triumph over this difficulty was achieved in \cite{DHR16} by exploiting the following observation:
The first order terms in \eqref{eq:Teuks} disappear for certain weighted higher-order derivatives of $\alpas$.
Indeed, if $\Ph$ denotes either $(\Omega^{-2}r^2\pu)^2(r\Omega^{4}\alpas[+2])$ or $(\Omega^{-2}r^2\pv)^2(r\alpas[-2])$, then \eqref{eq:Teuks} implies
\begin{equation}\label{eq:scat:RW}
\pu\pv \Ph=\frac{\Omega^2}{r^2}(\lap-4)\Ph+\frac{6M\Omega^2}{r^3}\Ph.
\end{equation}
Equation \eqref{eq:scat:RW}, known as the Regge--Wheeler equation, is nothing but the tensorial linear wave equation with a positive potential and, as such, can be treated using well-established methods.
In particular, \eqref{eq:scat:RW} admits a conserved energy: If $T_{\mu\nu}$ denotes the canonical energy momentum tensor associated to \eqref{eq:scat:RW}, then $\nabla^\mu(T_{\mu\nu}(\partial_t)^\nu)=0$.

To give context for the relevance of \eqref{eq:scat:RW} within \LG, let's summarise the strategy of the proof of stability of \LG~in~\cite{DHR16}.
First, the authors prove stability estimates for $\Ph$ using wave equation methods for \eqref{eq:scat:RW}. 
Then, they achieve from these estimates control over the extremal curvature components $\alpas[+2]$ and $\alpas[-2]$.
Finally, they use the control over $\alpas[+2]$ and $\alpas[-2]$, together with delicate considerations of gauge, to derive estimates for the entire remainder of \LG.

\subsection{Scattering theory for the Regge--Wheeler equation}
The construction of a scattering theory for \LG~follows a pattern similar to that of the proof of stability for~\LG: 
Given a seed scattering data set (Def.~\ref{def:seed}), we first derive expressions for the induced $\Ph=(\Omega^{-2}r^2\pu)^2(r\Omega^4\alpas[+2])$ along $\Cinn\cup\Scrim$. 

\begin{figure}[htpb]
\floatbox[{\capbeside\thisfloatsetup{capbesideposition={right,bottom},capbesidewidth=6cm}}]{figure}[\FBwidth]
{\caption{The limiting procedure that produces the scattering solution. }\label{fig:scat}}
{\includegraphics[width=120pt]{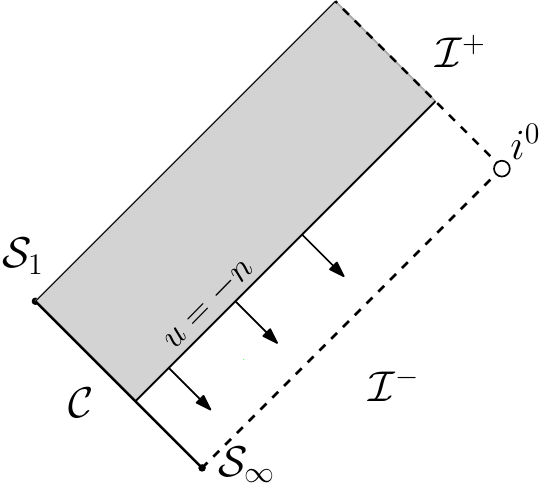}}
\end{figure}

We then temporarily ignore the rest of the system and only focus on the construction of the unique solution to \eqref{eq:scat:RW} restricting to these data.
This is done in the following way:
We consider a sequence of finite characteristic initial value problems (as depicted in Fig.~\ref{fig:scat}), where the data are posed on an outgoing null cone at finite retarded time $u=-n$ such that these data approach the scattering data along $\Scrim$ in the limit $n\to\infty$.
The unique existence of solutions to these finite characteristic initial value problems is ensured by standard theorems. 
We finally exploit the fact that \eqref{eq:scat:RW} admits a conserved energy to show uniform-in-$n$ estimates for these solutions, from which we deduce that the sequence of finite solutions constructed this way converges to a unique limiting solution---the scattering solution.
In fact, the conserved energy associated to \eqref{eq:scat:RW} defines \textit{unitary} Hilbert space isomorphisms between spaces of scattering data along $\Cinn\cup\Scrim$ and spaces of restrictions of the scattering solution to any Cauchy hypersurface in $D^+(\Cinn\cup\Scrim)$. For details, see \cite{Masaood22,V}.

\subsection{Scattering theory for the remainder of \texorpdfstring{\LG}{(LGS)}: }\label{sec:scat:remain}
At this point, we have a scattering solution $\Ph=(\Omega^{-2}r^2\pu)^2(r\Omega^4\alpas[+2])$ to \eqref{eq:scat:RW} that restricts correctly to the data along $\Cinn\cup\Scrim$ induced by the seed scattering data.
We can now construct the remaining quantities of \LG~by systemically defining other quantities in terms of the seed scattering data and the solution $\Ph$. 
For instance, by integrating $\Ph$ from $\Scrim$, and by computing the value of $\pu(r\Omega^4\alpas[+2])$ along $\Scrim$ from the seed scattering data, we can define $\pu(r\Omega^4\alpas[+2])$ in all of $D^+(\Cinn\cup\Scrim)$. Similarly, we define $\alpas[+2]$.

Owing to the high degree of redundancy in the system~\LG~(there are more equations than unknowns), this procedure is somewhat subtle. 
Let us illustrate the difficulty: The system \LG~includes equations for the $\pv-$, the $\pu-$ and angular derivatives for $\xh$, one of them being written down in \eqref{eq:math:exampleequations}. 
We could thus choose to define $\xh$ as the solution to the third of \eqref{eq:math:exampleequations} (assuming we have   already defined $\alpas[+2]$), but we might then have difficulties proving that the other equations for $\xh$ are also satisfied. 
Overcoming this problem requires being very careful about the order in which to define the various quantities and a precise understanding of the structure of the various equations.
See~\cite{Masaood22b,V} for details. 

We finish this section by remarking that the clean split that can be done for the linear problem, namely to first do a limiting argument to construct a solution $\Ph$ to \eqref{eq:scat:RW} and to then construct all the remaining quantities from this limiting solution $\Ph$, will likely be no longer possible in the full, nonlinear theory. 
Instead, one will have to construct a sequence of finite solutions to the entire system, and then show that this entire sequence converges.
\section{Solving the scattering problem II: Asymptotic analysis of solutions \texorpdfstring{$\alpas$}{alpha[s]} to the Teukolsky equations}\label{sec:asy1}
So far, we have written down scattering data along $\Cinn\cup\Scrim$ that capture the radiation emitted by an $N$-body system moving along approximately hyperbolic orbits in the infinite past (Def.~\ref{def:physicalseed}), and we have established that one can associate a unique scattering solution to these data (\S\ref{sec:scat}).
With this solution obtained, we shall now sketch how to obtain asymptotic properties of the solution, focussing first on the extremal Weyl components $\alpas[+2]$ and $\alpas[-2]$.  

\subsection{The induced scattering data for \texorpdfstring{$\alpas[+2]$}{alpha[+2]} and \texorpdfstring{$\alpas[-2]$}{alpha[s]}}\label{sec:asy1:induceddata}
Since both $\alpas[+2]$ and $\alpas[-2]$ satisfy decoupled equations, we can also decouple the presentation of their analysis from the rest of the system. 
For this, we merely need to write down the scattering data for $\alpas[+2]$ and $\alpas[-2]$ that are induced by seed scattering data as in Def.~\ref{def:physicalseed}.
We will henceforth use the notation 
\begin{equation}
r_0(u):= r(u,v=1)=|u|-2M\log|u|+\dots.
\end{equation}
\begin{lemma}
Given seed data satisfying Def.~\ref{def:physicalseed}, the induced data for $\alpas[+2]$ are given by:
\begin{align}\label{eq:al:data}
{\alpas[+2]_\ell}|_{\Cinn}=\aldata_\ell r_0^{-3}-\tfrac16 \tfrac{(\ell+2)!}{(\ell-2)!} \overline\albdata_\ell r_0^{-4}\log r_0+\mathscr{C}_\ell r_0^{-4}+\O(r_0^{-4-\epsilon}),&& \pv^n(r\alpas[+2])|_{\Scrim}=0,
\end{align}
for any $n\in\mathbb N_{\geq1}$.

Similarly, the induced data for $\alpas[-2]$ are given by 
\begin{align}\label{eq:alb:data}
\alpas[-2]|_{\Cinn}=\albdata_\ell r_0^{-4}+\O(r_0^{-4-\epsilon}),&& \pv^n(\Omega^{-2}r^2\pv)^2(r\alpas[-2])|_{\Scrim}=0\,\,\forall \,n\in\mathbb N_{\geq1}.
\end{align}
as well as the values on the spheres $\Sone$, $\Sinfty$ of the first two transversal derivatives:
\begin{align}\label{eq:alb:seeddata}
(\Omega^{-2}r^2\pv)(r\alpas[-2]_\ell)|_{\Sone}=\mathscr{C}'_{\ell},&& (\Omega^{-2}r^2\pv)^2(r\alpas[-2]_\ell)|_{\Sinfty}=2\overline{\aldata}_\ell.
\end{align}
Both $\mathscr{C}$ and $\mathscr{C}'$ are independent of $u$ and their precise values won't matter for this paper. 
\end{lemma}
The information from this lemma will suffice to derive the asymptotics for $\alpas[+2]$ and $\alpas[-2]$.
Notice that, even though $\alpas[+2]$ and $\alpas[-2]$ satisfy decoupled evolution equations, they are coupled at the level of the data. This is a manifestation of the Teukolsky--Starobinsky identities relating $\alpas[+2]$ and $\alpas[-2]$ \cite{Teukolsky74}.
\subsection{A sketch of the problem for the lowest angular mode \texorpdfstring{$\alpas[+2]_{\ell=2}$}{of alphas[2]}}\label{sec:asy1:l2}
Recall that the elliptic operator $\lap+s$ in \eqref{eq:Teuks} has eigenvalues $-\ell(\ell+1)+s(s+1)=-(\ell-s)(\ell+s+1)$, with $\ell\in\mathbb N_{\geq2}$.
In particular, for $\ell=s=2$, $(\lap+s)\alpas_{\ell}=0$, and \eqref{eq:Teuks} simplifies to:
\begin{equation}\label{eq:asy:Teukl=2}
\pu(r^{-4}\Omega^4\pv(r^5\alpas[2]_{\ell=2}))=-\frac{30M\Omega^6}{r^7}\cdot r^5\alpas[2]_{\ell=2}.
\end{equation}
The simple yet crucial observation now is that the RHS of \eqref{eq:asy:Teukl=2} has an extra decay of $r^{-3}$ compared to the LHS, even though the LHS only contains two derivatives.
Thus, if we have any preliminary global decay estimate of $\alpas$, then we can insert this estimate into the RHS of \eqref{eq:asy:Teukl=2} and integrate the equation in $u$ and $v$ to improve this estimate.

To illustrate this, let us assume that $|\alpas[2]_{\ell=2}|\lesssim r^{-p}$ for some $p>0$.
We then have that
$
|\pu(r^{-4}\Omega^4\pv(r^5\alpas[2]_{\ell=2}))|\lesssim M r^{-2-p}.
$
We now integrate this from $\Scrim$ in $u$, where $r^{-4}\pv(r^5\alpas[2]_{\ell=2})$ vanishes by \eqref{eq:al:data}. 
The result is that
$|\pv(r^5\alpas[2]_{\ell=2})|\lesssim M r^{3-p}$. 
Finally, we integrate this from $\Cinn$ in $v$ to obtain that $|r^5\alpas[2]_{\ell=2}-r^5\alpas[2]_{\ell=2}|_{\Cinn}|\leq M r^{4-p}$. 
Since $r^5\alpas[2]_{\ell=2}|_{\Cinn}$ is bounded by $u^2$ according to \eqref{eq:al:data}, we can divide by $r^5$ to improve the original estimate to $|\alpas[+2]_{\ell=2}|\lesssim r^{-5}u^2+r^{-p-1}$ and iterate to improve further.

In practice, the conservation of the energy associated to \eqref{eq:scat:RW} gives us the initial estimate $|\alpas[2]_{\ell=2}|\lesssim r^{-1}$, and iteratively inserting this estimate into \eqref{eq:asy:Teukl=2} gives us the estimate
\begin{equation}
|r^5\alpas_{\ell=2}-r^5\alpas_{\ell=2}|_{\Cinn}|\leq M\cdot r,\quad\implies\quad r^5\alpas_{\ell=2}=r^5\alpas_{\ell=2}|_{\Cinn}+\O(M\cdot r)=\aldata_{\ell=2}\cdot u^2+\dots.
\end{equation}
We now insert the last estimate into \eqref{eq:asy:Teukl=2} one last time, writing $\Omega^2=1-2M/r=1+\dots$:
\begin{equation}
\pv(r^5\alpas[2]_{\ell=2})=-30M\cdot r^4\int_{-\infty}^u \frac{\aldata_{\ell=2} u'^2}{r^7}+\dots \dd u'= -\frac{M\aldata_{\ell=2}}{2}+M\O(|u|/r),
\end{equation}
where the final equality follows from integrating by parts ($u^2/r^7=\frac{1}{6}\pu(u^2/r^6)+\frac{1}{15}\pu(|u|/r^5)+\frac{1}{15}1/r^5+\dots$; the last term being the dominant term near $\Scrip$).
By integrating again in $v$, we finally deduce that
\begin{equation}\label{eq:asy:al:l=2}
r^5\alpas_{\ell=2}=r^5\alpas_{\ell=2}|_{\Cinn}-\tfrac12M\cdot \aldata_{\ell=2} r+\dots=\aldata_{\ell=2} u^2-\tfrac12M\cdot \aldata_{\ell=2} r+\dots.
\end{equation}
Using this procedure, we can find higher and higher order terms. For instance, the next-to leading order term in $r^5\alpas[2]_{\ell=2}|_{\Cinn}$, namely $-4\overline{\albdata}_{\ell=2} r_0\log r_0$, adds the term $2M\overline{\albdata}_{\ell=2}\log^2 r$ to the expansion above.
This proves \eqref{eq:thm1} for $\ell=2$.
\subsection{A sketch for higher angular modes \texorpdfstring{$\alpas[+2]_{\ell}$}{of alpha[2]}}
\newcommand{\AN}[2][s]{{A}^{[#1],#2}}
For $\ell=2$, we crucially exploited the vanishing of $\lap+s$ in \eqref{eq:Teuks}. 
This made the RHS of \eqref{eq:Teuks} decay three powers faster in $r$ than the LHS, allowing us to iteratively improve any estimate that we insert into the RHS. 

For $\ell>2$, it looks like this cannot be done, since the RHS now only decays 2 powers faster, which no longer yields an improvement after integrating in $u$ and in $v$.
This hurdle is, similarly to how we descended from the Teukolsky equations to the Regge--Wheeler equations, overcome by commuting: Denote $\AN{N}:=(\Omega^{-2}r^2\pv)^{N}(r^{|s|+s+1}\alpas)$, and, abusing notation, $\lap\AN{N}:=(\Omega^{-2}r^2\pv)^{N}(r^{|s|+s+1}\lap\alpas)$. 
A  tedious computation, starting from \eqref{eq:Teuks}, gives
\begin{multline}\label{eq:asy:commuted}
\pu((\tfrac{\Omega^2}{r^2})^{N+s}\pv\AN{N})
=(\tfrac{\Omega^2}{r^2})^{N+s+1}(N(N + 1 + 2s)+(\lap + s))\AN{N}\\
+2M(\tfrac{\Omega^2}{r^2})^{N+s+1}\cdot (-c_N^{[s]}r^{-1}\AN{N}+ N(N + s)(N + 2s) \AN{N-1}),
\end{multline}
where $c_{N}^{[s]}=(1 + s)(1 + 2s) + 3N(N + 1 + 2s)$.
But if $\alpas$ is supported on angular frequency $\ell=N$, then the first line on the RHS of \eqref{eq:asy:commuted} vanishes, so we have
\begin{equation}\label{eq:asy:commutedLLL}
\pu((\tfrac{\Omega^2}{r^2})^{\ell}\pv\AN{\ell-s}_{\ell})
=2M(\tfrac{\Omega^2}{r^2})^{\ell+1}(-c_{\ell-s}^{[s]} r^{-1}\AN{\ell-s}_{\ell}+  (\ell\!-\!s)\ell(\ell\!+\!s) \AN{\ell-s-1}_{\ell}).
\end{equation}
The first term on the RHS now features an $r^{-3}$ weight multiplying $\AN{\ell-s}_{\ell}$, whereas the second term features an $r^{-2}$-weight that multiplies $\AN{\ell-s-1}_{\ell}$ instead of $\AN{\ell-s}_{\ell}$.
But since $\AN{N-1}=\AN{N-1}|_{\Cinn}+\int \frac{\Omega^2}{r^2}\AN{N}\dd v$, $\AN{N-1}$ decays one power faster compared to $\AN{N}$, so we can view this term to also feature three extra powers in decay compared to the LHS of \eqref{eq:asy:commutedLLL}.

At this point, for $s=2$, we can more or less mimic the argument presented in \S\ref{sec:asy1:l2} for $\ell=2$ to obtain an estimate for $\AN{\ell-s}_{\ell}$:

Firstly, we compute the transversal derivatives $\AN{N}_{\ell}|_{\Cinn}$, $N\leq \ell-s$, along $\Cinn$ by inductively integrating \eqref{eq:asy:commuted} from $\Scrim$, where for $s=2$ all boundary terms vanish .
The result of this is that $\AN{\ell-s}_{\ell}|_{\Cinn}=C_{\ell-s}\aldata_\ell r_0^{\ell}+\dots$ (and $\AN{N}_{\ell}|_{\Cinn}\sim r_0^{N+s}$ for $N\leq\ell-s$), with $C_{\ell-s}=\tfrac{(-1)^\ell 2(\ell-s)!(2\ell)!}{\ell!(\ell+s)!}$.

Secondly, we perform an iterative argument as for $\ell=2$ to produce a global estimate of the form
\begin{equation}\label{eq:asy:poop}
\pv\AN{\ell-s}_{\ell}=-M\aldata_{\ell}\frac{(-1)^\ell 2(\ell-s)!(\ell+1)!}{(\ell+s)!}r^{\ell-2}+M\O\left(\tfrac{|u|}{r}+\overline{\albdata}_{\ell}r^{\ell-3}\log r\right).
\end{equation} 
Thirdly and finally, we integrate this estimate $\ell-s+1$ times from $\Cinn$:
\begin{multline}\label{eq:asy:fundi}
r^{|s|+s+1}\alpas_{\ell}=\sum_{i=0}^{\ell-s} \frac{\AN{i}_{\ell}|_{\Cinn}}{i!}\left(\frac{1}{r_0}-\frac1r\right)^i \\
+\underbrace{\int_{v|_{\Cinn}}^v \frac{\Omega^2}{r^2}\cdots \int_{v|_{\Cinn}}^{v_{(\ell-s)}}\frac{\Omega^2}{r^2}}_{\text{$\ell-s+1$ integrals}}\left( \frac{r^2}{\Omega^2}\pv\AN{\ell-s}_{\ell}\right)\dd v_{(\ell-s+1)}\cdots \dd v_{1}.
\end{multline}
The computations involved are now getting more and more involved, but it is relatively straight-forward to see that the nested integral in the second line produces a leading-order term $(-1)^{\ell+1}\frac{2M\aldata_{\ell}r}{(\ell-1)(\ell+2)}$, whereas the sum in the first line clearly remains bounded as $v\to\infty$ along $u=\mathrm{const}$--this sum exactly corresponds to the sum in the first line in \eqref{eq:thm1}.
This proves~\eqref{eq:thm1}.
\subsection{A sketch for higher angular modes \texorpdfstring{$\alpas[-2]_{\ell}$}{of alpha[-2]}}\label{sec:asy1:alb}
Compared to $s=+2$, the main difference in the case $s=-2$ is that, with the decay rate $\alpas[-2]|_{\Cinn}\sim r_0^{-4}$, the RHS of \eqref{eq:Teuks} decays like $r^{-1}$, so we cannot integrate \eqref{eq:Teuks} from $\Scrim$ along $\Cinn$---instead, we have to integrate it from $\Sone$ in order to compute the transversal derivative $\pv(r\alpas[-s])$ along $\Cinn$.
This is why we had to specify $\pv\alpas[-2]$ along $\Sone$ as part of the seed data.

The commuted equation, \eqref{eq:asy:commuted} with $N=1$, can then be integrated from $\Scrim$, but the boundary term picked up from $\Scrim$, $\AN[s=-2]{2}_\ell|_{\Scrim}=2\overline{\aldata}_\ell\neq 0$ is a constant by \eqref{eq:alb:seeddata}.

At this point, all higher order transversal derivatives along $\Cinn$ can be computed as for $s=2$---all further boundary terms vanish at $\Scrim$ by \eqref{eq:alb:data}.
 In particular, we find that $\AN[s=-2]{\ell-s}_\ell|_{\Cinn}=\tfrac{(\ell-2)!}{(\ell+2)!}C_{\ell-s}\overline{\aldata}_\ell r_0^{\ell-s}+\dots$, so the leading order decay of higher-order transversal derivatives  along $\Cinn$  is not dictated by the leading order decay of $\alpas[-2]|_{\Cinn}$ itself (i.e.~it does not depend on~$\albdata$)!

We can now, in exact analogy to \eqref{eq:asy:poop}, show that
\begin{equation}
\pv\AN[-2]{\ell-s}_{\ell}=(-1)^{\ell+1} M\overline{\aldata}_{\ell}\cdot(\ell+1)!r^{\ell-2}+\dots
\end{equation}
from which we derive an expression for $\alpas[-2]$ by again applying the formula \eqref{eq:asy:fundi}:
\begin{equation}
r\alpas[-2]_{\ell}=\sum_{i=0}^{\ell+2} \frac{1}{i!}\left(\frac{1}{r_0}-\frac1r\right)^i \AN{i}_{\ell}|_{\Cinn}+2M\overline{\aldata}_{\ell}\frac{(-1)^{\ell+1}\cdot \ell(\ell+1)}{6r^3}\log (\tfrac{r}{r_0})+\dots.
\end{equation}
The leading-order logarithmic behaviour is entirely governed by $\overline{\aldata}$, instead of the coefficient $\albdata$ that governs the leading order decay of $\alpas[-2]$ towards $\Scrim$, the reason being that this behaviour is determined by higher-order transversal derivatives along $\Cinn$.\footnote{This important detail was overlooked in the brief outlook given in \cite{GK}, which therefore made an incorrect prediction on $s=-2$.}

However, if we compute the expression for the radiation field of $\alpas[-2]_\ell$, namely $\lim_{v\to\infty}r\alpas[-2]_{\ell}$, the coefficient $\overline\aldata$ only enters at Schwarzschildean order, that is to say: It does not contribute if $M=0$. 
A very lengthy computation produces the second statement of Theorem~\ref{thmIV}, \eqref{eq:thm2}. See~\cite{V} for details.

\subsection{The summing of the \texorpdfstring{$\ell$-}{angular }modes}\label{sec:asy1:summing}
The method we used to find the asymptotics for higher angular modes comes at a high price: Commuting with $(\Omega^{-2}r^2\pv)$ up to $\ell-s$ times generates a large amount of error terms that, although they all decay faster towards infinity (i.e.~in $1/r$), individually grow extremely fast in $\ell$. 
In particular, if we write $\alpas=\sum_{\ell=2}^{\infty}\alpas_{\ell}$, the bounds for the error terms in the estimates above are not summable even if the initial data are smooth.

We expect that this is only an artefact of the method that we use, and that there is a way to sum the estimates above. The investigation of this expectation is ongoing work~\cite{X}. Very roughly speaking, we hope to tackle the issue in four steps:
\begin{enumerate}[label=(\arabic*)]
\item Consider the initial data for $\alpas[2]$: $\alpas[2]|_{\Cinn}=\aldata r_0^{-3}+\albdata r_0^{-4}\log r_0+\mathscr{C} r_0^{-4}+\O(r_0^{-4-\epsilon})$. We write the solution $\alpas[2]$ as a sum of the solution $\alpas[2]_{\mathrm{phg}}$ arising from the initial data $\alpas[2]_{\mathrm{phg}}|_{\Cinn}=\aldata r_0^{-3}+\albdata r_0^{-4}\log r_0+\mathscr{C} r_0^{-4}$ (without the $\O$-term) and the difference $\alpas[2]_{\Delta}$. 
\item A robust energy estimate on the difference $\alpas[2]_{\Delta}$ then proves that $\alpas[2]_{\Delta}=\O(r^{-4-\epsilon})$ in $D^+(\Cinn\cup\Scrim)$.
\item  Next, we consider $\alpas[2]_{\mathrm{phg}}$. For this solution, we establish a persistence of polyhomogeneity result. Let's explain this:
Loosely speaking, a function is said to be polyhomogeneous near a boundary $x=\infty$ if it has a series expansion in $x^{-p}\log^q x$, with a countable set of numbers $p\in\mathbb R$, $q\in\mathbb N$, in a robust, square-integrable sense. 
What we will show is that, since the initial data for $\alpas[2]_{\mathrm{phg}}$ are polyhomogeneous near $\Scrim$, $\alpas[s]_{\mathrm{phg}}$ will also be polyhomogeneous near $i^0$ and near $\Scrip$. This complements the result of~\cite{HV20}.
\item We can finally determine the coefficients of the polyhomogeneous expansion of $\alpas[+2]$ by decomposing $\alpas[2]_{\mathrm{phg}}$ into angular frequencies and using the methods of Section~\ref{sec:asy1}---since we have already established that $\alpas[2]_{\mathrm{phg}}$ is polyhomogeneous, we no longer need to worry about issues of summability at this stage.
\end{enumerate}

\section{Solving the scattering problem III: Asymptotic analysis of the remainder of \texorpdfstring{\LG}{(LGS)}}\label{sec:asy2}
Equipped with asymptotic expressions for $\alpas[+2]$ and $\alpas[-2]$, we are now in a position to derive the asymptotic behaviour of all other quantities contained in \LG.
For instance, asymptotic expressions for $\xh$, $\xhb$ and $\gsh$ follow from integrating the equations \eqref{eq:math:exampleequations}. 
To obtain the asymptotic behaviour for the remaining Weyl curvature quantities $\overone{\Psi}_i$, $i=1,2,3$, one needs to integrate the Bianchi identities; the result of this is that $\alpas[+2]\sim r^{-4}$ near $\Scrip$ (recall that $\alpas[+2]$ is equivalent to $\Ps_0$) implies that $\Ps_1$, denoted $\be$ in \cite{DHR16}, goes like $r^{-4}\log r$ near $\Scrip$ (as predicted by \cite{Chr02}), whereas the peeling rates \eqref{eq:intro:peeling} hold true for $\Ps_2$, $\Ps_3$ and $\Ps_4$.

In order to obtain asymptotic expressions for all the other quantities contained in \LG, we would of course have to introduce \LG~in much more detail.
For the scope of the current paper, suffice it to say that the constructed solution can be proved to be \emph{extendable to $\Scrip$ for any $s<1$} in the sense of Def.~3.4 of \cite{Holz16} (which also features a compact introduction of \LG). 
We now move on to highlighting a few other points of particular interest.

\subsection{Bondi normalising \texorpdfstring{$\Scrip$}{Scri-plus} in addition to \texorpdfstring{$\Scrim$}{Scri-minus}}
Provided we have Bondi normalised our seed scattering data at $\Scrim$, i.e.~made the metric perturbations $\Om$ and $\gs$, as well as $r^2\K$ and $r\trx$ vanish at $\Scrim$, it is natural to ask whether we can Bondi normalise the resulting solution along $\Scrip$ as well.
The answer is affirmative:
Without going into too much detail, starting from Bondi normalised seed scattering data and our results for $\alpas[+2]$ and $\alpas[-2]$, we derive that the limits $r^2\K|_{\Scrip}$ and $\gsh|_{\Scrip}$ vanish automatically. 
On the other hand, we find that the limits $\Om|_{\Scrip}$ and $r\trxb|_{\Scrip}$ exist and decay in $u$. 
It is now possible to add a pure gauge solution that simultaneously kills off both of these terms. Since this gauge solution decays in $u$, it does not affect the Bondi normalisation of $\Scrim$. 
\subsection{Corrections to the quadrupole formula: \texorpdfstring{$\dd E/\dd u$ along $\Scrip$}{Bondi mass loss along Scri-plus}}
In the non-linear theory, the rate of change of energy along $\Scrip$ is given by the Bondi mass loss formula~\cite{Bondi60Nature}: $\frac{\dd E}{\dd u}=-\frac14\langle|\bm r \bm{\underline{\hat{\chi}}}|_{\Scrip}|^2\rangle$, $\langle \cdot\rangle$ denoting the average over $\Stwo$ and $\bm r \bm{\underline{\hat{\chi}}}|_{\Scrip}$ denoting the News function at $\mathcal{I}^+$. 
Of course, the linearisation of this formula is zero, but we can still define the linearised energy loss to be
\begin{equation}
\frac{\overone{\dd{E}}}{\dd u}:=-\frac1{16\pi} \int_{\Stwo}|r\xhb|_{\Scrip}|^2 \sin\theta\dd\theta\dd\varphi.
\end{equation}
An expression for $r\xhb$ can be inferred via integration of the second of \eqref{eq:math:exampleequations} and our asymptotic expression \eqref{eq:thm2} for $\alpas[-2]$:
\begin{equation}\label{eq:asy2:quadr}
\lim_{u=\mathrm{const},v\to\infty}r\xhb_\ell=(-1)^{\ell+1}|u|^{-3}\left(\frac{\albdata_{\ell}}{\ell(\ell+1)}+\frac{2M\overline{\aldata}_{\ell}(\ell-2)!}{(\ell+2)!}\right)+\O(|u|^{-3-\epsilon}).
\end{equation}
For $\ell=2$ and $M=0$, this can be seen to exactly coincide with Einstein's original quadrupole formula, though perhaps in a slightly less familiar form: Equation~\eqref{eq:asy2:quadr} relates the limiting behaviour near $\Scrim$ (governed by $\aldata$ and $\albdata$ according to Def.~\ref{def:physicalseed})  to the limiting behaviour along~$\Scrip$ as $u\to-\infty$. 
The full relation to the quadrupole formula is then established by relating the coefficients $\aldata_{\ell=2}$ and $\albdata_{\ell=2}$ to the Newtonian quadrupole moments \eqref{eq:phys:Newtonmoments} as described in \S\ref{sec:phys}.

More generally, \eqref{eq:asy2:quadr} represents the linear corrections to (the multipole generalisation of) this formula when the linearisation is done around Schwarzschild instead of Minkowski.\footnote{In parts of the literature, such corrections arising from backscattering off of the curvature of the spacetime are referred to as ``tail effects'', see \cite{Blanchet93, Blanchet23} and references therein.}
Since the present work is meant as a precursor to the treatment of the full Einstein vacuum equations, we hope to eventually make the propagation part of this formula, i.e.~going from $\Scrim$ to $\Scrip$, precise. 
Establishing a full, non-linear relation between the coefficients $\aldata$ and $\albdata$ and the physical multipole moments of the described matter distribution would be an exciting project on its own, cf.~footnote~\ref{fottynote}. 

\subsection{The antipodal matching condition}
Since the analysis produces an arbitrarily precise understanding of the asymptotic behaviour of solutions near $i^0$, it is also relevant for several questions concerning relations between the past limit point of $\Scrip$ and the future limit point of $\Scrim$. 
To give an example, we can prove the following about the magnetic parts of the News functions:
\begin{equation}
\lim_{v\to\infty}\lim_{u\to-\infty}r^2{\xhb}^{\mathrm{B2}}_{\ell}=(-1)^{\ell}\lim_{u\to-\infty}\lim_{v\to\infty}r^2{\xh}^{\mathrm{B2}}_{\ell},
\end{equation}
This is closely related to the antipodal matching condition. See also~\cite{Strominger14,Masaood22b,CNP22}.
\section{The question of late-time asymptotics}\label{sec:late}
The entirety of the previous sections was concerned with the asymptotic behaviour near infinity restricted to $u\leq -1$. 
It turns out, however, that this, in a certain sense, entirely fixes the possible asymptotics at late times, i.e.~as $u\to\infty$, due to the existence of certain conserved, modified Newman--Penrose charges along $\Scrip$.
This is discussed in detail in \cite{GK} at the level of the scalar wave equation, together with a heuristic principle to relate solutions of the Teukolsky equation to scalar waves (\S6 of \cite{GK}).
Unfortunately, a small mistake crept into the latter heuristics: 
In the equation above (6.1), it was assumed that the global asymptotic behaviour of $\Psi_4$ (a.k.a. $\alpas[-2]$--we now drop the $\overone{}$) would be governed by its decay towards $\Scrim$, so $p_4$ was taken to be 3. 
But as we stressed in \S\ref{sec:asy1:alb} of the present paper, it is the decay of two transversal derivatives of $\Psi_4$ near~$\Scrim$ that determines the logarithmic behaviour near $\Scrip$, and the value of $p_4$ should instead be $2$.

\begin{figure}[htpb]
\floatbox[{\capbeside\thisfloatsetup{capbesideposition={right,bottom},capbesidewidth=6cm}}]{figure}[\FBwidth]
{\caption{We can extend the scattering data along $\C$ all the way to the future event horizon in order to obtain the asymptotics near future timelike infinity. These turn out to be completely fixed by the asymptotics at early times, i.e., they are independent of how we extend the data along $\C$ to the future. 
We note here that, if the spacetime were stationary in the past, then the late-time asymptotics would be three powers faster \cite{GK,MZ21}.}\label{fig:late}}
{\includegraphics[width=180pt]{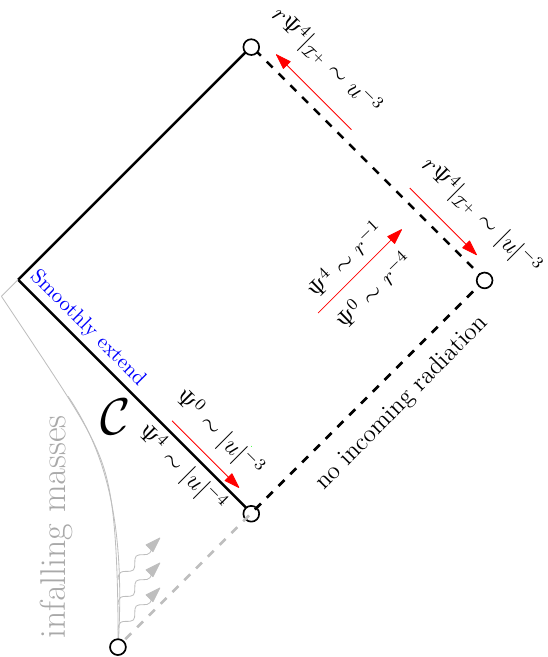}}
\end{figure}

Thus, if we extend, for instance, the data along $\C$ all the way to the future event horizon, we end up with the global asymptotics  depicted in  Fig.~\ref{fig:late}. See also \S1.6 of \cite{V}.
\section{Conclusion}\label{sec:conclusion}
Clearly, the results \eqref{eq:thm1} and \eqref{eq:asy:al:l=2}, which show that $\alpas[+2]\sim r^{-4}$ near $\Scrip$ (in complete agreement with~\cite{Damour86}), violate the peeling property \eqref{eq:intro:peeling} for $i=0$ (recall that $\alpas[+2]$ is equivalent to $\Psi_0$). 
As opposed to the violation of peeling near $\Scrim$, which is derived using Post-Newtonian approximations around Minkowski (\S\ref{sec:phys}), the violation of peeling near $\Scrip$ is caused by the mass term in the Schwarzschild metric (in the Post--Minkowskian picture, this effect would only be seen at order $\epsilon^2$ in $g=\eta+\epsilon g^1+\epsilon^2 g^2+\O(\epsilon^3)$).

While our results thus rule out that the constructed spacetimes admit a smooth $\Scrip$ (or $\Scrim$) or admit Bondi coordinates near $\Scrip$, both of these concepts can still be made sense of provided they are sufficiently weakened:
In the case of Bondi coordinates, it suffices to drop condition (iii) from \cite{SeriesVIII}. 
To be precise, the $b$-term in (4.1) of \cite{SeriesVIII}, which captures the $1/r^2$ fall-off of $\gamma$ and $\delta$ in (2.9) of \cite{SeriesVIII},  does \textit{not} vanish, it carries physical information. See \cite{Kroon99} for the computation of these terms if $\Psi_0\sim r^{-4}$. 
The non-vanishing of this term then generates a $r^{-3}\log r$ in the expansion of $U$, a $r^{-1}\log r$ term in the expansion of $V=-r+2M+\dots$, etc. See also \S1.5.1 of \cite{V}.

In the case of Penrose's asymptotic simplicity, the assumption on the regularity of the compactification needs to be weakened significantly. 
A nontrivial problem of interest to many would be to understand the optimal regularity with which a conformal boundary can be still be attached if $\Psi^0=\O(r^{-4})$.\footnote{Based on the results of \cite{HV20} (see Remark~8.2 therein), one may expect this regularity to be $C^{1,\alpha}$ for any $\alpha<1$. Note, however, that not even the question of the minimal conformal regularity that ensures \eqref{eq:intro:peeling} is resolved, cf.~\cite{Friedrich02} and \S2 of \cite{Friedrich18}.}

Another notion of asymptotic flatness mentioned in the introduction is that of \cite{CK93}, which demands the spacetime to feature certain decay towards spatial infinity (along $t=0$).
The demanded decay in particular implies that $\alpas[+2]$ would decay like $o(r^{-\frac{7}{2}})$ towards $i^0$. Now, inspection of \eqref{eq:thm1} or \eqref{eq:asy:Teukl=2} immediately shows that $\alpas[+2]\sim r^{-3}$ along $t=0$. In other words, the class of spacetimes constructed in \cite{CK93} decays too fast to capture the kinds of physics we are describing in this paper.
Furthermore, we can see that the interpretation of Christodoulou's argument~\cite{Chr02} given in (the paragraphs above equation (2.4)) of \cite{I} does not hold, at least at the linearised level:  For, if we set the coefficient $\aldata$ to vanish in Def.~\ref{def:physicalseed}, this will neither affect the no incoming radiation condition on $\Scrim$, nor the decay rate of the energy loss along $\Scrip$, see eq.~\eqref{eq:asy2:quadr}. 
Moreover, the induced data along $t=0$ will decay fast enough for the framework of \cite{CK93} to apply.
But at the same time, $\alpas[+2]$ will now decay like $r^{-5}\log^2 r$ towards $\Scrip$ by \eqref{eq:thm1}, from which one easily shows that $\be$ decays like $r^{-4}$. But this is in contradiction with the argument, which states that $\be\sim r^{-4}\log r$; cf.~(1.10) of~\cite{I}. 
See \S1.5.4 of \cite{V} for an extended discussion.

Finally, a notion of asymptotic flatness that does admit our constructed spacetimes is that of~\cite{DHR16}. 
Indeed, as we said in \S\ref{sec:asy2}, the constructed spacetime is ``extendable to $\Scrip$'' in the sense of Def.~3.4 of \cite{Holz16} for any~$s<1$; the parameter $s$ measuring decay towards $\Scrip$. 

The fact that our constructions decay half a power slower near $i^0$ than demanded by \cite{CK93} but still decay sufficiently fast near $\Scrip$ is somewhat surprising as the spacetimes constructed in \cite{CK93} are generically  only ``extendable to $\Scrip$'' for $s\leq 1/2$. 
This remarkable improvement in decay towards~$\Scrip$ compared to decay near $i^0$ is related to our constructions not having incoming radiation from~$\Scrim$ and 
will be further discussed elsewhere.
\small{\bibliographystyle{alpha} 
\bibliography{references_all}}

\end{document}